\documentclass[12pt]{iopart}
\usepackage{pdflscape}
\usepackage{epsfig}
\usepackage{iopams}

\renewcommand{\theequation}{\thesection.\arabic{equation}}
\newcommand{\be}{\begin{equation}}
\newcommand{\ee}{\end{equation}}

\newcommand{\ba}{\begin{array}}
\newcommand{\ea}{\end{array}}
\newcommand{\bea}{\begin{eqnarray}}
\newcommand{\eea}{\end{eqnarray}}

\newcommand{\sdir}{\ensuremath{\rlap{\raisebox{.15ex}{$\mskip
6.5mu\scriptstyle+ $}}\supset}}

\newtheorem{proposition}{Proposition}

\begin{document}

\title[SUSY versions of the equations of conformally parametrized surfaces]{Supersymmetric versions of the equations of conformally parametrized surfaces}
\author{S Bertrand$^1$, A M Grundland$^{2,3}$ and A J Hariton$^2$}

\address{$^1$ Department of Mathematics and Statistics, Universit\'e de Montr\'eal,\\ Montr\'eal CP 6128 Succ. Centre-Ville (QC) H3C 3J7, Canada}
\address{$^2$ Centre de Recherches Math\'ematiques, Universit\'e de Montr\'eal,\\ Montr\'eal CP 6128 Succ. Centre-Ville (QC) H3C 3J7, Canada}
\address{$^3$ Department of Mathematics and Computer Science, Universit\'e du Qu\'ebec, Trois-Rivi\`eres, CP 500 (QC) G9A 5H7, Canada}
\ead{bertrans@crm.umontreal.ca, grundlan@crm.umontreal.ca and hariton@crm.umontreal.ca}

\begin{abstract} The objective of this paper is to formulate two distinct supersymmetric (SUSY) extensions of the Gauss-Weingarten and Gauss-Codazzi (GC) equations for conformally parametrized surfaces immersed in a Grassmann superspace, one in terms of a bosonic superfield and the other in terms of a fermionic superfield. We perform this analysis using a superspace-superfield formalism together with a SUSY version of a moving frame on a surface. In constrast with the classical case, where we have three GC equations, we obtain six such equations in the bosonic SUSY case and four such equations in the fermionic SUSY case. In the fermionic case the GC equations resemble the form of the classical GC equations. We determine the Lie symmetry algebra of the classical GC equations to be infinite-dimensional and perform a subalgebra classification of the one-dimensional subalgebras of its largest finite-dimensional subalgebra. We then compute superalgebras of Lie point symmetries of the bosonic and fermionic SUSY GC equations respectively, and classify the one-dimensional subalgebras of each superalgebra into conjugacy classes. We then use the symmetry reduction method to find invariants, orbits and reduced systems for two one-dimensional subalgebras for the classical case, two one-dimensional subalgebras for the bosonic SUSY case and two one-dimensional subalgebras for the fermionic SUSY case. We find explicit solutions of these reduced SUSY systems, which correspond to different surfaces immersed in a Grassmann superspace. Within this framework for the SUSY versions of the GC equations, a geometrical interpretation of the results is discussed.

\paragraph{}Keywords: supersymmetric models, Lie superalgebras, symmetry reduction, conformally parametrized surfaces.
\end{abstract}


\maketitle

\section{Introduction}
In the last three decades, a number of supersymmetric (SUSY) extensions of classical and quantum mechanical models, describing several physical phenomena, have been developed and group-invariant solutions of these SUSY systems have been found (e.g. \cite{BJ01}-\cite{JP00}). Recently, this method was further generalized to encompass hydrodynamic-type systems (see e.g. \cite{Das}-\cite{Hariton}). Their SUSY extensions were established and their group-invariant solutions were constructed. SUSY versions of the Chaplygin gas in (1+1) and (2+1) dimensions were formulated by R. Jackiw \textit{et al}, derived from parametrizations of the action for a superstring and a Nambu-Goto membrane respectively (see \cite{Jackiw} and references therein). It was suggested that a quark-gluon plasma may be described by non-Abelian fluid mechanics \cite{TJZW}. In addition, SUSY extensions have been formulated for a number of integrable equations , including among others the Korteweg-de Vries equation \cite{Chaichian}-\cite{Mathieu}, the Kadomtsev-Petviashvili equation \cite{Manin}, the Sawada-Kotera equation \cite{Tian} and the sine-Gordon and sinh-Gordon equations \cite{Aratyn}-\cite{GHS09}. Various approaches have been used to construct supersoliton solutions, such as the inverse scattering method, B\"acklund and Darboux transformations for odd and even superfields, Lax formalism in a superspace and generalized versions of the symmetry reduction method (SRM). A number of supersoliton and multi-supersoliton solutions were determined by a Crum-type transformation \cite{Grammaticos},\cite{Siddiq05},\cite{Matveev} and it was found that there exist infinitely many local conserved quantities. A connection was established between the super-Darboux transformations and super-B\"acklund transformations which allow for the construction of supersoliton solutions \cite{Chaichian},\cite{Liu},\cite{Tian},\cite{Aratyn},\cite{Gomes},\cite{Siddiq06},\cite{GHS09}.

Despite the progress made in the investigation of nonlinear SUSY systems, this area of mathematics does not yet have as solid a theoretical foundation as the classical theory of differential equations. 
This is related primarily to the fact that, due to the nature of Grassmann variables, the principle of superposition of solutions obtained from the method of characteristics cannot be applied to nonlinear SUSY systems.
In most cases, analytic methods for solving quasilinear SUSY systems of equations lead to the construction of classes of solutions that are more restricted than the general solution. One can attempt to construct more restricted classes of solutions which depend on some arbitrary functions and parameters by requiring that the solutions be invariant under certain group transformations of the original system. The main advantages of the group properties appear when group analysis makes it possible to construct regular algorithms for finding certain classes of solutions without referring to any additional considerations but proceeding directly from the given system of partial differential equations (PDEs). A systematic computational method for constructing the group of symmetries of a given system of PDEs has been developed by many authors (see e.g. G. W. Bluman and S. C. Anco \cite{Bluman}, P. A. Clarkson and P. Winternitz \cite{Clarkson}, P. Olver \cite{Olver}, and D. Sattinger and O. Weaver \cite{SW}) and a broad review of recent developments in the SUSY theory can be found in several books (e.g. J. F. Cornwell \cite{Cornwell}, B. DeWitt \cite{DeWitt}, D. S. Freed \cite{Freed}, V. Kac \cite{Kac} and V. S. Varadarajan \cite{Varadarajan}). The methodological approach adopted in this paper is based on the use of the SRM to find solutions of the PDEs which are invariant under subgroups of a Lie supergroup of point transformations. By a symmetry supergroup of a SUSY system of PDEs, we mean a local SUSY Lie supergroup $G$ transforming both the independent and dependent variables of the considered SUSY system of equations in such a way that $G$ transforms given solutions of the system to new solutions of the same system. The Lie superalgebra of such a supergroup is represented by vector fields and their prolongation structures. The standard algorithms for determining the symmetry algebra of a system of equations and classifying its subalgebras have been extended in order to deal with SUSY models (see e.g. \cite{Kac},\cite{RH90},\cite{Winternitz}).

Recent studies of the geometric properties of surfaces associated with holomorphic and non-holomorphic solutions of the SUSY bosonic Grassmann sigma models have been performed \cite{Delisle}-\cite{Witten}. A gauge-invariant formulation of these SUSY models in terms of orthogonal projectors allows one to obtain explicit solutions and consequently to study the geometry of their associated surfaces. In differential geometry, parametrized surfaces are described in terms of a moving frame satisfying the Gauss-Weingarten (GW) equations, which are linear PDEs. Their compatibility conditions are the Gauss-Codazzi (GC) equations. A representation of nonlinear equations in the form of the GC equations is the starting point in the theory of integrable (soliton) surfaces arising from infinitesimal deformations of integrable differential equations and describing the behaviour of soliton solutions. The construction and analysis of such surfaces associated with integrable systems in several areas of mathematical physics provide new tools for the investigation of nonlinear phenomena described by these systems. In this setting, it is our objective to perform a systematic analysis of SUSY versions of the GW and GC equations. The formulation of a SUSY extension of the GW and GC equations has already been accomplished for the specific case of bosonic Grassmann sigma models \cite{Delisle}. It is of considerable interest to consider such extensions for the general case of the GW and GC equations.

The purpose of this paper is to formulate two distinct SUSY extensions of the GW and GC equations, one using a bosonic superfield and the other using a fermionic superfield, for conformally parametrized surfaces in the superspace $\mathbb{R}^{(n_b\vert n_f)}$. The SUSY versions of these equations are formulated through the use of a superspace-superfield formalism. The considered surfaces are parametrized by the vector field $\mathcal{F}$ and the normal vector field $\mathcal{N}$, which are replaced in the SUSY version by their corresponding superfields $F$ and $N$ in $\mathbb{R}^{(n_b\vert n_f)}$. This allows us to formulate the SUSY extensions of the structural equations for the immersion of conformally parametrized surfaces explicitly in matrix form. We establish explicit forms of the SUSY GW equations satisfied by the moving frame on these surfaces. The result is independent of the parametrization. This allows us to examine various geometric properties of the studied immersions, such as the first and second fundamental forms of the surfaces (and therefore the mean and Gaussian curvatures).

The paper is organized as follows. The symmetry algebra of the classical GC equations is determined and a subalgebra classification of its one-dimensional subalgebras is performed in section 2. In section 3, we introduce the basic properties of Grassmann algebras and Grassmann variables and introduce the notation that will be used in what follows. In section 4, we construct the bosonic and fermionic SUSY extensions of the GW and GC equations. In section 5, we discuss certain geometric aspects of the conformally parametrized SUSY surfaces. We provide expressions for the first and second fundamental forms and the Gaussian and mean curvatures, which are required for a geometrical interpretation of the invariant solutions. In section 6, we determine Lie superalgebras of point symmetries of the SUSY GC equations for both the bosonic and fermionic cases. Section 7 involves a classification of the one-dimensional subalgebras of both Lie superalgebras into conjugacy classes. In section 8, we provide examples of invariant solutions of the supersymmetric Gauss-Codazzi equations obtained by the SRM. Finally, in section 9, we present the conclusions and discuss possible future developments in this field.

\section{Symmetries of conformally parametrized surfaces}
The system of PDEs describing the moving frame $\Omega=(\partial \mathcal{F},\bar{\partial}\mathcal{F},\mathcal{N})^T$ on a smooth conformally parametrized surface in $3$-dimensional Euclidean space satisfies the following GW equations
\begin{equation}
\partial\Omega=V_1\Omega,\qquad\bar{\partial}\Omega=V_2\Omega,\label{ClaGW}
\end{equation}
where the matrices $V_1$ and $V_2$ are given by
\begin{equation}
\hspace{-1.5cm}V_1=\left(\begin{array}{ccc}
\partial u&0&Q\\
0&0&\frac{1}{2}He^u\\
-H&-2Qe^{-u}&0
\end{array}\right),\qquad V_2=\left(\begin{array}{ccc}
0&0&\frac{1}{2}He^u\\
0&\bar{\partial}u&\bar{Q}\\
-2\bar{Q}e^{-u}&-H&0
\end{array}\right).\label{ClaV1V2}
\end{equation}
Here $\partial$ and $\bar{\partial}$ are the partial derivatives with respect to the complex coordinates $z=x+iy$ and $\bar{z}=x-iy$, respectively. The conformal parametrization of a surface is given by a vector-valued function $\mathcal{F}=(\mathcal{F}_1,\mathcal{F}_2,\mathcal{F}_3):\mathcal{R}\rightarrow\mathbb{R}^3$ (where $\mathcal{R}$ is a Riemann surface) which satisfies the following normalization for the tangent vectors $\partial \mathcal{F}$ and $\bar{\partial}\mathcal{F}$ and the unit normal $\mathcal{N}$
\begin{equation}
\begin{array}{lll}
\langle\partial \mathcal{F},\partial \mathcal{F}\rangle=\langle\bar{\partial}\mathcal{F},\bar{\partial}\mathcal{F}\rangle=0,& &\langle\partial \mathcal{F},\bar{\partial}\mathcal{F}\rangle=\frac{1}{2}e^u,\\
\langle\partial \mathcal{F},\mathcal{N}\rangle=\langle\bar{\partial}\mathcal{F},\mathcal{N}\rangle=0,& &\langle \mathcal{N},\mathcal{N}\rangle=1.
\end{array}
\end{equation}
We define the quantities
\begin{equation}
Q=\langle\partial^2\mathcal{F},\mathcal{N}\rangle\in\mathbb{C},\qquad H=2e^{-u}\langle\partial\bar{\partial}\mathcal{F},\mathcal{N}\rangle\in\mathbb{R},
\end{equation}
where $Qdz^2$ and $\bar{Q}d\bar{z}^2$ are the Hopf differentials and $H$ is the mean curvature function. Here, the bracket $\langle\cdot,\cdot\rangle$ denotes the scalar product in 3-dimensional Euclidean space $\mathbb{R}^3$
\begin{equation}
\langle a,b\rangle=a_1b_1+a_2b_2+a_3b_3.\label{innerproduct}
\end{equation}
So, the GW equations for a moving frame $\Omega$ on a surface have to obey the following system of equations
\begin{equation}
\hspace{-2cm}\begin{array}{l}
\partial^2\mathcal{F}=\partial u\partial \mathcal{F}+Q\mathcal{N},\qquad \partial\bar{\partial}\mathcal{F}=\frac{1}{2}He^u\mathcal{N},\qquad\bar{\partial}^2\mathcal{F}=\bar{\partial}u\bar{\partial}\mathcal{F}+\bar{Q}\mathcal{N},\\
\partial \mathcal{N}=-H\partial \mathcal{F}-2e^{-u}Q\bar{\partial}\mathcal{F},\qquad \bar{\partial}\mathcal{N}=-2e^{-u}\bar{Q}\partial \mathcal{F}-H\bar{\partial}\mathcal{F}.
\end{array}
\end{equation}
The first and second fundamental forms are given by
\begin{equation}
\hspace{-2.5cm}I=\langle d\mathcal{F},d\mathcal{F}\rangle=\left\langle\frac{e^u}{2}\left(\begin{array}{cc}
0&1\\
1&0
\end{array}\right)\left(\begin{array}{c}
dz\\
d\bar{z}
\end{array}\right),\left(\begin{array}{c}
dz\\
d\bar{z}
\end{array}\right)\right\rangle=e^u\left\langle\left(\begin{array}{c}
dx\\
dy
\end{array}\right),\left(\begin{array}{c}
dx\\
dy
\end{array}\right)\right\rangle,\label{ClaI}
\end{equation}
and
\begin{equation}
\hspace{-2.5cm}I\hspace{-0.1cm}I=\langle d^2\mathcal{F},\mathcal{N}\rangle=\left\langle\left(\begin{array}{cc}
Q+\bar{Q}+e^uH&i(Q-\bar{Q})\\
i(Q-\bar{Q})&-(Q+\bar{Q})+e^uH
\end{array}\right)\left(\begin{array}{c}
dx\\
dy
\end{array}\right),\left(\begin{array}{c}
dx\\
dy
\end{array}\right)\right\rangle,\label{ClaII}
\end{equation}
respectively. The principal curvatures $k_1$ and $k_2$ are the eigenvalues of the matrix
\begin{equation}
B=e^{-u}\left(\begin{array}{cc}
Q+\bar{Q}+e^uH&i(Q-\bar{Q})\\
i(Q-\bar{Q})&-(Q+\bar{Q})+e^uH
\end{array}\right).
\end{equation}
We obtain the following expressions for the mean and Gaussian curvatures
\begin{equation}
H=\frac{1}{2}(k_1+k_2)=\frac{1}{2}\mbox{tr}(B),
\end{equation}
\begin{equation}
\mathcal{K}=k_1k_2=\det(B)=H^2-4Q\bar{Q}e^{-2u}.
\end{equation}
Umbilic points on a surface take place when $H^2-\mathcal{K}=0$ which implies that $\vert Q\vert^2=0$. The compatibility conditions of the GW equations (\ref{ClaGW}) are the GC equations
\begin{equation}
\bar{\partial}V_1-\partial V_2+[V_1,V_2]=0,
\end{equation}
(the bracket $[\cdot,\cdot]$ denotes the commutator) which reduce to the following three differential equations for the quantities $Q$, $H$ and $e^u$
\begin{equation}
\hspace{-1cm}\begin{array}{cr}
\partial\bar{\partial}u+\frac{1}{2}H^2e^u-2Q\bar{Q}e^{-u}=0,&\mbox{(the Gauss equation)}\\
\partial \bar{Q}-\frac{1}{2}e^u\bar{\partial}H=0,\qquad\bar{\partial}Q-\frac{1}{2}e^u\partial H=0&\mbox{(the Codazzi equations).}
\end{array}
\label{ClaGC}
\end{equation}
These equations are the necessary and sufficient conditions for the existence of conformally parametrized surfaces in 3-dimensional Euclidean space $\mathbb{R}^3$ with the fundamental forms given by (\ref{ClaI}) and (\ref{ClaII}). A review of systematic computational methods for constructing surfaces for a given moving frame can be found in several books (e.g. \cite{Baer}-\cite{Thorpe}). Equations (\ref{ClaGW}), (\ref{ClaV1V2}) and (\ref{ClaGC}) allow us to formulate explicitly the structural equations for the immersion directly in matrix terms. However, it is non-trivial to identify those surfaces which have an invariant geometrical characterization \cite{do Carmo},\cite{Bob}. The task of finding an increasing number of solutions of the GW and GC equations is related to the group properties of these systems of equations. The methodological approach adopted here is based on the SRM for PDEs invariant under a Lie group $G$ of point transformations.
Using the Maple program, we find that the symmetry group of the classical GC equations (\ref{ClaGC}) consists of conformal scaling transformations. The corresponding symmetry algebra $\mathcal{L}_1$ is spanned by the vector fields
\be
\ba{l}
X(\eta)=\eta(z)\partial_z+\eta'(z)(-2Q\partial_Q-U\partial_U),\\
Y(\zeta)=\zeta(\bar{z})\partial_{\bar{z}}+\zeta'(\bar{z})(-2\bar{Q}\partial_{\bar{Q}}-U\partial_U),\\
e_0=-H\partial_H+Q\partial_Q+\bar{Q}\partial_{\bar{Q}}+2U\partial_U,
\ea\label{ClaSymGC}
\ee
where $\eta'(\cdot)$ and $\zeta'(\cdot)$ are the derivatives of $\eta(\cdot)$ and $\zeta(\cdot)$ with respect to their arguments and where we have used the notation $e^u=U$. The commutation relations are
\begin{equation}
\hspace{-2.5cm}\begin{array}{l}
[X(\eta_1),X(\eta_2)]=(\eta_1\eta'_2-\eta'_1\eta_2)\partial_z+(\eta_1''\eta_2-\eta_1\eta_2'')(2Q\partial_Q+U\partial_U),\\
\left[Y(\zeta_1),Y(\zeta_2)\right]=(\zeta_1\zeta'_2-\zeta'_1\zeta_2)\partial_{\bar{z}}+(\zeta_1''\zeta_2-\zeta_1\zeta_2'')(2\bar{Q}\partial_{\bar{Q}}+U\partial_U),\\
\left[ X(\eta),Y(\zeta) \right]=0,\qquad [X(\eta),e_0]=0, \qquad [Y(\zeta),e_0]=0.
\end{array}
\end{equation}
Since the vector fields $X(\eta), Y(\zeta)$ and $e_0$ form an Abelian algebra, they determine that the algebra $\mathcal{L}_1$ can be decomposed as a direct sum of two infinite-dimensional Lie algebras together with a one-dimensional algebra generated by $e_0$, i.e.
\be
\mathcal{L}_1=\lbrace X(\eta)\rbrace\oplus\lbrace Y(\zeta)\rbrace\oplus\lbrace e_0\rbrace.
\ee
This algebra represents a direct sum of two copies of the Virasoro algebra together with the one-dimensional subalgebra $\lbrace e_0\rbrace$. Assuming that the functions $\eta$ and $\zeta$ are analytic in some open subset $\mathcal{D}\subset\mathbb{C}$, we can develop $\eta$ and $\zeta$ as power series with respect to their arguments and provide a basis for $\mathcal{L}_1$. The largest finite-dimensional subalgebra $L_1$ of the algebra $\mathcal{L}_1$ is spanned by the seven generators
\be
\hspace{-1.5cm}\ba{l}
e_0=-H\partial_H+Q\partial_Q+\bar{Q}\partial_{\bar{Q}}+2U\partial_U,\\
e_1=\partial_z,\qquad e_3=z\partial_z-2Q\partial_Q-U\partial_U,\qquad e_5=z^2\partial_z-4zQ\partial_Q-2zU\partial_U,\\
e_2=\partial_{\bar{z}},\qquad e_4=\bar{z}\partial_{\bar{z}}-2\bar{Q}\partial_{\bar{Q}}-U\partial_U,\qquad e_6=\bar{z}^2\partial_{\bar{z}}-4\bar{z}\bar{Q}\partial_{\bar{Q}}-2\bar{z}U\partial_U,
\ea\label{ClaSymFGC}
\ee
with non-zero commutation relations
\be
\ba{lll}
\phantom{~}[e_1,e_3]=e_1,& [e_1,e_5]=2e_3,& [e_3,e_5]=e_5,\\
\phantom{~}[e_2,e_4]=e_2,& [e_2,e_6]=2e_4,& [e_4,e_6]=e_6.
\ea
\ee
This seven-dimensional Lie subalgebra $L_1$ can be decomposed as a direct sum of two simple subalgebras together with a one-dimensional algebra generated by $e_0$,
\be
L_1=\lbrace e_1, e_3, e_5\rbrace\oplus\lbrace e_2,e_4, e_6\rbrace\oplus\lbrace e_0\rbrace.\label{ClaDecGC}
\ee
Therefore, the classification of the subalgebras of $L_1$ consists of two copies of a 3-dimensional Lie algebra together with the center $\lbrace e_0\rbrace$. This 3-dimensional Lie algebra corresponds to the algebra $A_{3,8}$ in the classification of J. Patera and P. Winternitz \cite{Patera} which is isomorphic to $\mathfrak{su}(1,1)$. The resulting classification of the subalgebras of $L_1$ into conjugacy classes, performed according to the methods described in \cite{Winternitz}, is given by the following list of representative subalgebras $L_{1,j}$
\be
\hspace{-2.7cm}\ba{lll}
L_{1,0}=\lbrace e_0\rbrace, & \hspace{-1cm}L_{1,1}=\lbrace e_1\rbrace, & \hspace{-2cm}L_{1,2}=\lbrace e_3\rbrace, \qquad L_{1,3}=\lbrace e_1+e_5\rbrace,\hspace{-2cm} \\
L_{1,4}=\lbrace e_2\rbrace, & L_{1,5}=\lbrace e_4\rbrace, & L_{1,6}=\lbrace e_2+e_6\rbrace,  \\
L_{1,7}=\lbrace e_1+\epsilon e_2\rbrace, & L_{1,8}=\lbrace e_1+\epsilon e_4\rbrace, & L_{1,9}=\lbrace e_2+e_6+\epsilon e_1\rbrace, \\
L_{1,10}=\lbrace e_3+\epsilon e_2\rbrace, & L_{1,11}=\lbrace e_3+ae_4\rbrace, & L_{1,12}=\lbrace e_2+e_6+ae_3\rbrace,  \\
L_{1,13}=\lbrace e_1+e_5+\epsilon e_2\rbrace, & L_{1,14}=\lbrace e_1+e_5+ae_4\rbrace, & L_{1,15}=\lbrace e_1+e_5+a(e_2+e_6)\rbrace,  \hspace{-2cm}
\ea\label{Cla1D}
\ee
where $\epsilon=\pm1$ and $a\neq0$ are parameters. The center of $L_1$, $\lbrace e_0\rbrace$, can be added to any of the subalgebras given above, say $L_{1,j}=\lbrace A\rbrace$, to produce a twisted subalgebra of the form $L_{1,j}'=\lbrace A+be_0\rbrace$, where $b\neq0$. The symmetry reductions associated with the subalgebras (\ref{Cla1D}) lead to systems of ordinary differential equations (ODEs). These reduced systems were analyzed systematically as a single generic symmetry reduction in \cite{ConteGrundland}, where the GC equations (\ref{ClaGC}) were reduced to the most general Painlev\'e P6 form (containing two or three arbitrary parameters).

In the following sections, the symmetry properties of the classical GW and GC equations are compared with their bosonic and fermionic SUSY counterparts.

\section{Preliminaries on Grassmann algebras}\setcounter{equation}{0}
The mathematical background formalism is based on the theory of supermanifolds as presented in \cite{Cornwell}-\cite{Varadarajan},\cite{Berezin}-\cite{Weinberg} and can be summarized as follows. The starting point in our consideration is a complex Grassmann algebra $\Lambda$ involving a finite or infinite number of Grassmann generators $(\xi_1,\xi_2,...)$. The number of Grassmann generators of $\Lambda$ is not essential provided that there is a sufficient number of them to make any formula encountered meaningful. The Grassmann algebra $\Lambda$ can be decomposed into its even and odd parts
\begin{equation}
\Lambda=\Lambda_{even}+\Lambda_{odd},
\end{equation}
where $\Lambda_{even}$ includes all terms involving a product of an even number of generators $\xi_k$, i.e. $1,\xi_1\xi_2,\xi_1\xi_3,...$, while $\Lambda_{odd}$ includes all terms involving a product of an odd number of generators $\xi_k$, i.e. $\xi_1,\xi_2,\xi_3,...,\xi_1\xi_2\xi_3,...$ The elements of $\Lambda$ are called supernumbers. The even supernumbers, variables, fields, etc are assumed to be elements of the even part $\Lambda_{even}$ of the underlying abstract real (complex) Grassmann ring $\Lambda$. The odd supernumbers, variables, fields, etc lie in its odd part $\Lambda_{odd}$. 
In the context of supersymmetry, the spaces $\Lambda$ and/or $\Lambda_{even}$ replace the field of complex numbers.  The Grassmann algebra can also be decomposed as
\begin{equation}
\Lambda=\Lambda_{body}+\Lambda_{soul},
\end{equation}
where
\begin{equation}
\Lambda_{body}=\Lambda^0[\xi_1,\xi_2,...]\simeq\mathbb{C},\qquad\Lambda_{soul}=\sum_{k\geqslant1}\Lambda^k[\xi_1,\xi_2,...].
\end{equation}
Here $\Lambda^0[\xi_1,\xi_2,...]$ refers to all terms that do not involve any of the generators $\xi_i$, while $\Lambda^k[\xi_1,\xi_2,...]$ refers to all terms that involve products of $k$ generators (for instance, if we have 4 generators $\xi_1,\xi_2,\xi_3,\xi_4$, then $\Lambda^2[\xi_1,\xi_2,\xi_3,\xi_4]$ refers to all terms involving $\xi_1\xi_2$, $\xi_1\xi_3$, $\xi_1\xi_4$, $\xi_2\xi_3$, $\xi_2\xi_4$ or $\xi_3\xi_4$). The bodiless elements in $\Lambda_{soul}$ are non-invertible because of the $\mathbb{Z}_0^+$-grading of the Grassmann algebra. If the number of Grassmann generators $\mathfrak{K}$ is finite, bodiless elements are nilpotent of degree at most $\mathfrak{K}$. In this paper, we assume that $\mathfrak{K}$ is arbitrarily large but finite. Our analysis is based on the global theory of supermanifolds as described in \cite{BerezinMono}-\cite{Rogers80}.

\paragraph{}Next, in our consideration, we use a $\mathbb{Z}_2$-graded complex vector space $V$ which has even basis elements $u_i$, $i=1,2,...,N$, and odd basis elements $v_\mu$, $\mu=1,2,...,N$, and construct $W=\Lambda\otimes_\mathbb{C}V$. We are interested in the even part of $W$
\begin{equation}
W_{even}=\left\lbrace\ba{c|c}
\sum_ia_iu_i+\sum_\mu\underline{\alpha}_\mu v_\mu & a_i\in\Lambda_{even}, \underline{\alpha}_\mu\in\Lambda_{odd}
\ea\right\rbrace.
\end{equation}
Clearly, $W_{even}$ is a $\Lambda_{even}$ module which can be identified with $\Lambda^{\times N}_{even}\times\Lambda_{odd}^{\times M}$ (consisting of $N$ copies of $\Lambda_{even}$ and $M$ copies of $\Lambda_{odd}$). We associate with the original basis, consisting of $u_i$ and $v_\mu$ (although $v_\mu\in\hspace{-0.35cm}\setminus\hspace{0.2cm} W_{even}$), the corresponding functionals
\bea
E_j: W_{even}\rightarrow\Lambda_{even}: E_j\left(\sum_ia_iu_i+\sum_\mu\underline{\alpha}_\mu v_\mu\right)=a_j,\\
\Upsilon_\nu: W_{even}\rightarrow\Lambda_{odd}: \Upsilon_\nu\left(\sum_ia_iu_i+\sum_\mu\underline{\alpha_\mu} v_\mu\right)=\underline{\alpha}_\nu,
\eea
and view them as the coordinates (even and odd respectively) on $W_{even}$. Any topological manifold locally diffeomorphic to a suitable $W_{even}$ is called a supermanifold.

\paragraph{}The transitions to even and odd coordinates between different charts on the supermanifold are assumed to be even- and odd-valued superanalytic or at least $G^\infty$ functions on $W_{even}$. A comprehensive definition of the classes of supersmooth functions $G^\infty$ and superanalytic functions $G^\omega$ can be found in \cite{Rogers80}, definition 2.5. We note that superanalytic functions are those that can be expanded into a convergent power series in even and odd coordinates, whereas the definition of the $G^\infty$ function is a more involved analogue of functions on manifolds. Any $G^\infty$ function can be expanded into products of odd coordinates in a Taylor-like expansion but the coefficients, being functions of even and odd coordinates, may not necessarily be analytic (see e.g. \cite{Rogers80}).

\paragraph{}The super-Minkowski space can be viewed as such a supermanifold globally diffeomorphic to $\Lambda_{even}^{\times 2}\times \Lambda_{odd}^{\times 2}$ with even light-cone coordinates $x_+,x_-$ and odd coordinates $\theta^+,\theta^-$. Here $x_+$ and $x_-$ are linear combinations of terms involving an even number of generators : $1,\xi_1\xi_2,\xi_1\xi_3,\xi_1\xi_4,...,\xi_2\xi_3,\xi_2\xi_4,...,\xi_1\xi_2\xi_3\xi_4,...$ On the other hand, $\theta^+$ and $\theta^-$ are linear combinations of terms involving an odd number of generators : $\xi_1,\xi_2,\xi_3,\xi_4,...,\xi_1\xi_2\xi_3,\xi_1\xi_2\xi_4,\xi_1\xi_3\xi_4,\xi_2\xi_3\xi_4,...$ The SUSY transformation (\ref{SUSYTrans1}) and (\ref{SUSYTrans2}) in the next section can be viewed as a particular change of coordinates on $\mathbb{R}^{(1,1\vert2)}$ which transforms solutions of the SUSY GW equations (\ref{SUSYGW}) and SUSY GC equations, (\ref{SUSYZCC}) respectively, into solutions of the same systems in new coordinates. A smooth superfield is a $G^\infty$ function from $\mathbb{R}^{(n_b\vert n_f)}$ to $\Lambda$ (where $n_b$ and $n_f$ are the numbers of bosonic and fermionic coordinates, respectively). It can be expanded in powers of the odd coordinates $\theta^+$ and $\theta^-$ giving a decomposition in terms of even superfields
\begin{equation*}
\chi_{even}:\Lambda_{even}^{\times2}\rightarrow\Lambda_{even},
\end{equation*}
and odd superfields
\begin{equation*}
\chi_{odd}:\Lambda_{even}^{\times2}\rightarrow\Lambda_{odd}.
\end{equation*}

\paragraph{}In this paper, we use the convention that partial derivatives involving odd variables satisfy the Leibniz rule
\begin{equation}
\partial_{\theta^\pm}(hg)=(\partial_{\theta^\pm}h)g+(-1)^{\deg(h)}h(\partial_{\theta^\pm}g),
\end{equation}
where the degree of a homogeneous supernumber is given by
\begin{equation}
\deg(h)=\left\lbrace\begin{array}{c}
0\mbox{ if }h\mbox{ is even,}\\
1\mbox{ if }h\mbox{ is odd,}
\end{array}\right.
\end{equation}
and we have used the notation
\begin{equation}
f_{\theta^+\theta^-}=\partial_{\theta^-}\left(\partial_{\theta^+}f\right).
\end{equation}
The partial derivatives with respect to the odd coordinates satisfy $\partial_{\theta^i}\theta^j=\delta_i^{\phantom{i}j}$ where the indices $i$ and $j$ each stand for $+$ or $-$ and $\delta_i^{\phantom{i}j}$ is the Kronecker delta function. The operators $\partial_{\theta^\pm}$, $J_\pm$ and $D_\pm$, in equations (\ref{D}) and (\ref{J}) change the parity of a bosonic function to a fermionic function and vice versa. For example, if $\phi$ is an even function, then $\partial_{\theta^+}\phi$ is an odd superfield while $\partial_{\theta^+}\partial_{\theta^-}\phi$ is an even superfield and so on. The chain rule for a Grassmann-valued composite function $f(g(x_+))$ is
\begin{equation}
\frac{\partial f}{\partial x_+}=\frac{\partial g}{\partial x_+}\frac{\partial f}{\partial g}.
\end{equation}
The interchange of mixed derivatives (with proper respect to the ordering of odd variables) is assumed throughout. For further details see e.g. the books by Cornwell \cite{Cornwell},  DeWitt \cite{DeWitt}, Freed \cite{Freed}, Varadarajan \cite{Varadarajan} and references therein.

\section{SUSY extensions of the GW and GC equations}\setcounter{equation}{0}
The purpose of this section is to establish two different SUSY extensions of the GW and GC equations, one using a bosonic superfield representation and the other using a fermionic superfield representation of a surface in a superspace ($\mathbb{R}^{(2,1\vert2)}$ for the bosonic extension and $\mathbb{R}^{(1,1\vert3)}$ for the fermionic extension). Let $\mathcal{S}$ be a smooth simply connected surface in a Minkowski superspace with the bosonic light-cone coordinates $x_+=\frac{1}{2}(t+x)$ and $x_-=\frac{1}{2}(t-x)$ together with the fermionic (anti-commuting) variables $\theta^+$ and $\theta^-$ such that
\begin{equation}
(\theta^+)^2=(\theta^-)^2=\theta^+\theta^-+\theta^-\theta^+=0.\label{theta2}
\end{equation}
We assume that the surface $\mathcal{S}$ is conformally parametrized by a  vector-valued superfield $F(x_+,x_-,\theta^+,\theta^-)$ (bosonic in the case of the bosonic SUSY extension, fermionic in the case of the fermionic SUSY extension,) which can be decomposed as
\begin{equation}
\hspace{-2.5cm}F=F_m(x_+,x_-)+\theta^+\varphi_m(x_+,x_-)+\theta^-\psi_m(x_+,x_-)+\theta^+\theta^-G_m(x_+,x_-),\quad m=1,2,3\label{F}
\end{equation}
For the bosonic SUSY extensions, the functions $F_m$ and $G_m$ are bosonic (even Grassmann)-valued, while the functions $\varphi_m$ and $\psi_m$ are fermionic (odd Grassmann)-valued. Conversely, for the fermionic SUSY extension, the functions $F_m$ and $G_m$ are fermionic-valued, while the functions $\varphi_m$ and $\psi_m$ are bosonic-valued.
Here, the fields $F_m$, $\varphi_m$, $\psi_m$ and $G_m$ are the four parts of the truncated power series with respect to $\theta^+$ and $\theta^-$ of the $m^{th}$ component of the vector superfield $F$. Power series with respect to $\theta^+$ and $\theta^-$ are truncated since the fermionic variables $\theta^+$ and $\theta^-$ satisfy (\ref{theta2}). Also, let $D_+$ and $D_-$ be the covariant superspace derivatives
\be
D_\pm=\partial_{\theta^\pm}-i\theta^\pm\partial_{x_\pm}.\label{D}
\ee
The covariant derivatives $D_\pm$ have the property that they anticommute with the differential SUSY operators
\begin{equation}
J_+=\partial_{\theta^+}+i\theta^+\partial_{x_+},\qquad J_-=\partial_{\theta^-}+i\theta^-\partial_{x_-},\label{J}
\end{equation}
which generate the SUSY transformations
\begin{equation}
x_+\rightarrow x_++i\underline{\eta}_+\theta^+,\qquad \theta^+\rightarrow\theta^++i\underline{\eta}_+,\label{SUSYTrans1}
\end{equation}
and
\begin{equation}
x_-\rightarrow x_-+i\underline{\eta}_-\theta^-,\qquad \theta^-\rightarrow\theta^-+i\underline{\eta}_-,\label{SUSYTrans2}
\end{equation}
respectively. Here $\underline{\eta}_+$ and $\underline{\eta}_-$ are odd-valued parameters. The four operators, $D_+$, $D_-$, $J_+$ and $J_-$ satisfy the anticommutation relations
\begin{equation}
\hspace{-2.5cm}\lbrace J_n,J_m\rbrace=2i\delta_{mn}\partial_{x_m},\quad \lbrace D_m,D_n\rbrace=-2i\delta_{mn}\partial_{x_m},\quad \lbrace J_m,D_n\rbrace=0,\quad m,n=1,2\label{DJ}
\end{equation}
where $\delta_{ij}$ is the Kronecker delta function and $\lbrace\cdot,\cdot\rbrace$ denotes the anticommutator, unless otherwise noted. Here, the values $1$ and $2$ of the indices $m$ and $n$ stand for $+$ and $-$, respectively. Therefore we have the following relations
\begin{equation}
D_\pm^2=-i\partial_\pm,\qquad J_\pm^2=i\partial_\pm.\label{D2}
\end{equation}

The conformal parametrization of the surface $\mathcal{S}$ in the superspace $\mathbb{R}^{(n_b\vert n_f)}$ is assumed to give the following normalization of the superfield $F$
\be
\langle D_iF,D_jF\rangle=g_{ij}f,\qquad i,j=1,2.\label{NormalizationF}
\ee
Here the values $1$ and $2$ of the indices $i$ and $j$ stand for $+$ and $-$, respectively. The scalar product $\langle\cdot,\cdot\rangle$ in (\ref{NormalizationF}) is defined in the same way as in equation (\ref{innerproduct}), taking into account the property (\ref{theta2}) regarding the odd-valued variables $\theta^+$ and $\theta^-$, and taking values in the Grassmann algebra $\Lambda$.
Hence the bosonic functions $g_{ij}$ of $x_+$, $x_-$, $\theta^+$ and $\theta^-$ are given by
\be
g_{11}=0,\qquad g_{12}=\frac{1}{2}e^\phi,\qquad g_{21}=\frac{\epsilon}{2}e^\phi,\qquad g_{22}=0,\label{gij}
\ee
where $\epsilon=-1$ in the bosonic case and $\epsilon=1$ in the fermionic case. Therefore, in the bosonic case, $f$ is a bodiless bosonic function (i.e. $f\in\Lambda_{soul}$) of $x_+$ and $x_-$ which is nilpotent of order $k$. In the fermionic case, $f$ is a bosonic function which may or may not be bodiless. In the bosonic case, the bodiless function $f(x_+,x_-)$ has been introduced since the normalization $\langle D_+F,D_-F\rangle$ contains only terms with products of generators $\xi_i$ and the exponential contains a term which involves no generator. One should note that the equations (\ref{NormalizationF}) are identically satisfied for $i=j$ in the bosonic extension, which is not the case in the fermionic extension. Indeed, in the scalar product (\ref{innerproduct}), we have the sum of the squares of each $m^{th}$ component of the tangent vector superfield $D_iF$. Since the square of a fermionic function vanishes, each of the terms in the scalar product is identically zero, i.e. 
\be
\langle D_iF,D_iF\rangle=0.
\ee
In the case of the mixed scalar product, the normalization condition is
\be
\langle D_+F,D_-F\rangle=\frac{1}{2}e^\phi f.
\ee
It is interesting to note that, by construction, the metric coefficients $g_{ij}$ of the bosonic extension are antisymmetric for $i\neq j$, i.e.
\be
g_{ij}=-g_{ji}.\label{Bgij}
\ee
This is in contrast with the fermionic case where the coefficients of the induced metric $g_{ij}$ are symmetric in the indices $i$ and $j$, i.e.
\be
g_{ij}=g_{ji}.\label{Fgij}
\ee

The superfield $\phi$ is assumed to be bosonic and can be decomposed as the following power series in the fermionic variables $\theta^+$ and $\theta^-$
\begin{equation}
\phi =u(x_+,x_-)+\theta^+\gamma(x_+,x_-)+\theta^-\delta(x_+,x_-)+\theta^+\theta^-v(x_+,x_-),\label{phi}
\end{equation}
where $u$ and $v$ are bosonic-valued functions, while $\gamma$ and $\delta$ are fermionic-valued functions. Through a power expansion in $\theta^+$ and $\theta^-$ we find the exponential form
\begin{equation}
\begin{array}{l}
e^\phi=e^{u}(1+\theta^+\gamma+\theta^-\delta+\theta^+\theta^-(v-\gamma\delta)),\\
e^{-\phi}=e^{-u}(1-\theta^+\gamma-\theta^-\delta-\theta^+\theta^-(v+\gamma\delta)).
\end{array}
\end{equation}

The tangent vector superfields $D_\pm F$ together with the normal bosonic superfield $N(x_+,x_-,\theta^+,\theta^-)$, which can be decomposed as
\begin{equation}
\hspace{-2.5cm}N=N_m(x_+,x_-)\hspace{-0.1cm}+\hspace{-0.1cm}\theta^+\alpha_m(x_+,x_-)\hspace{-0.1cm}+\hspace{-0.1cm} \theta^-\beta_m(x_+,x_-)\hspace{-0.1cm}+\hspace{-0.1cm}\theta^+\theta^-H_m(x_+,x_-),\quad m=1,2,3\label{N}
\end{equation}
form a moving frame $\Omega$ on the surface $\mathcal{S}$ in the superspace. Here, the bosonic-valued fields $N_m$ and $H_m$ and the fermionic-valued fields $\alpha_m$ and $\beta_m$ are the four parts of the truncated power series with respect to $\theta^+$ and $\theta^-$ of the $m^{th}$ component of the vector superfield $N$. This normal superfield $N$ has to satisfy the conditions
\begin{equation}
\langle D_iF,N\rangle=0,\qquad\langle N,N\rangle=1,\qquad i=1,2.\label{NormalizationN}
\end{equation}

We now derive the bosonic and fermionic SUSY versions of the GW and GC equations.
We assume that we can decompose the second-order covariant derivatives of $F$ and first-order derivatives of $N$ in terms of the tangent vectors $D_+F$ and $D_-F$ and the unit normal $N$,
\begin{equation}
\begin{array}{l}
D_jD_iF=\Gamma_{ij}^{\phantom{ij}k}D_kF+b_{ij}fN,\\
D_iN=b^k_{\phantom{k}i}D_kF+\omega_i N,
\end{array}\qquad i,j,k=1,2\label{SUSYdec}
\end{equation}
where the coefficients $\omega_i$ and $\Gamma_{ij}^{\phantom{ij}k}$ are fermionic functions. However, the functions $b_{ij}$ and $b^k_{\phantom{k}i}$ are bosonic-valued in the bosonic extension and are fermionic-valued in the fermionic extension. The SUSY GW equations for the moving frame $\Omega$ on a surface are given by
\begin{equation}
D_+\Omega=A_+\Omega,\qquad D_-\Omega=A_-\Omega,\qquad\Omega=\left(\ba{c}
D_+F\\
D_-F\\
N
\ea\right),\label{SUSYGW}
\end{equation}
where the $3\times3$ supermatrices $A_+$ and $A_-$ are
\be
A_+=\left(\ba{ccc}
\Gamma_{11}^{\phantom{11}1} & \Gamma_{11}^{\phantom{11}2} & b_{11}f \\
\Gamma_{21}^{\phantom{21}1} & \Gamma_{21}^{\phantom{21}2} & b_{21}f \\
b^1_{\phantom{1}1} & b^2_{\phantom{2}1} & \omega_1
\ea\right),\qquad A_-=\left(\ba{ccc}
\Gamma_{12}^{\phantom{12}1} & \Gamma_{12}^{\phantom{12}2} & b_{12}f \\
\Gamma_{22}^{\phantom{22}1} & \Gamma_{22}^{\phantom{22}2} & b_{22}f \\
b^1_{\phantom{1}2} & b^2_{\phantom{2}2} & \omega_2
\ea\right).\label{SUSYA}
\ee
The conformally parametrized surface $\mathcal{S}$ satisfies the normalization conditions (\ref{NormalizationF}) and (\ref{NormalizationN}) for the superfields $F$ and $N$, and we define the quantities $Q^+,Q^-$ and $H$ to be
\be
b_{11}=Q^+,\qquad b_{12}=\frac{1}{2}e^\phi H,\qquad b_{21}=-\frac{1}{2}e^\phi H, \qquad b_{22}=Q^-,\label{QH}
\ee
which gives the relations
\be
\hspace{-1.5cm}\langle D_+^2F,N\rangle=Q^+f,\qquad \langle D_-D_+F,N\rangle=\frac{1}{2}e^\phi Hf,\qquad \langle D_-^2F,N\rangle=Q^-f.\label{QHnorm}
\ee
The coefficients of the second fundamental form $b_{ij}$ have the property
\be
b_{ij}=-b_{ji},\qquad\mbox{for}\qquad i\neq j.\label{antibij}
\ee
To obtain a relation between the functions $b_{ij}$ and $b^k_{\phantom{k}j}$, we make use of the relation
\be
\langle D_jD_iF,N\rangle=D_j\langle D_iF,N\rangle-\epsilon\langle D_iF,D_jN\rangle=-\epsilon\langle D_iF,D_jN\rangle,
\ee
and by substituting $D_jN$ into its decomposition (\ref{SUSYdec}) we get the relation
\be
\left(g_{ik}b^k_{\phantom{k}j}+\epsilon b_{ij}\right)f=0.\label{SUSYrelbij}
\ee
We can obtain the coefficients $\omega_i$ from the derivative of $\langle N,N\rangle=1$, i.e.
\be
0=D_i\langle N,N\rangle=\langle D_iN,N\rangle+\langle N,D_iN\rangle=2\omega_i\langle N,N\rangle=2\omega_i,
\ee
from which we obtain
\begin{equation}
\omega_i=0.
\end{equation}
Also, we can make use of the identities
\begin{equation}
\hspace{-1.5cm}\ba{l}
D_k(\frac{1}{2}e^\phi f)=D_k\langle D_+F,D_-F\rangle=\langle D_kD_+F,D_-F\rangle -\langle D_+F,D_kD_-F\rangle\\
\phantom{D_k(\frac{1}{2}e^\phi f)}=\Gamma_{1k}^{\phantom{1k}1}\langle D_+F,D_-F\rangle +\Gamma_{2k}^{\phantom{2k}2}\langle D_+F,D_-F\rangle,
\ea
\end{equation}
which lead to
\begin{equation}
D_kf=(\Gamma_{1k}^{\phantom{1k}1}+\Gamma_{2k}^{\phantom{2k}2}-D_k\phi)f.\label{Df}
\end{equation}
From equation (\ref{Df}) we can compute the compatibility condition on the function $f$
\be
\lbrace D_+,D_-\rbrace f=\left(D_-\Gamma_{11}^{\phantom{11}1}+D_-\Gamma_{21}^{\phantom{21}2}+D_+\Gamma_{12}^{\phantom{12}1}+ D_+\Gamma_{22}^{\phantom{22}2}\right)f=0.\label{DDf}
\ee
Hence we define the Christoffel symbols of the first kind $\Gamma_{ijk}$ to be
\be
\Gamma_{ijk}f=\langle D_jD_iF,D_kF\rangle.
\ee
By construction, the Christoffel symbols of the first and second kind ($\Gamma_{ijk}$ and $\Gamma_{ij}^{\phantom{ij}k}$, respectively) are antisymmetric under permutation of the indices $i$ and $j$, i.e.
\be
\Gamma_{ijk}=-\Gamma_{jik},\qquad \Gamma_{ij}^{\phantom{ij}k}=-\Gamma_{ji}^{\phantom{ij}k},\quad\mbox{for}\quad i\neq j.\label{antiChrist}
\ee
The relations between the Christoffel symbols of first and second kind are given by
\be
\Gamma_{ijk}f=\langle D_iD_jF,D_kF\rangle=\Gamma_{ij}^{\phantom{ij}l}\langle D_lF,D_kF\rangle
\ee
or
\be
(\Gamma_{ijk}-\Gamma_{ij}^{\phantom{ij}l}g_{lk})f=0.
\ee

The compatibility conditions of the SUSY GW equations are the SUSY GC equations, given by
\begin{equation*}
\hspace{-2.5cm}\ba{l}
\lbrace D_+,D_-\rbrace\Omega=D_+(A_-\Omega)+D_-(A_+\Omega),\\
\phantom{\lbrace D_+,D_-\rbrace\Omega}=D_+A_-\Omega+\left(\ba{ccc}
-\Gamma_{11}^{\phantom{11}1} & -\Gamma_{11}^{\phantom{11}2} & -\epsilon b_{11}f \\
\Gamma_{12}^{\phantom{12}1} & \Gamma_{12}^{\phantom{12}2} & -\epsilon b_{21}f \\
-\epsilon b^1_{\phantom{1}1} & -\epsilon b^2_{\phantom{2}1} & 0
\ea\right)D_+\Omega\\
\\
\phantom{\lbrace D_+,D_-\rbrace\Omega}+D_-A_+\Omega+\left(\ba{ccc}
-\Gamma_{12}^{\phantom{12}1} & -\Gamma_{12}^{\phantom{12}2} & -\epsilon b_{12}f \\
-\Gamma_{22}^{\phantom{22}1} & -\Gamma_{22}^{\phantom{22}2} & -\epsilon b_{22}f \\
-\epsilon b^1_{\phantom{1}2} & -\epsilon b^2_{\phantom{2}2}  & 0
\ea\right)D_-\Omega\\
\phantom{\lbrace D_+,D_-\rbrace\Omega}=D_+A_-\Omega-EA_-ED_+\Omega+D_-A_+\Omega-EA_+ED_-\Omega.
\ea
\end{equation*}
So we have
\be
D_+A_-+D_-A_+-\lbrace EA_+,EA_-\rbrace=0,\label{SUSYZCC}
\ee
up to the addition of a non-zero matrix $P$ such that $P\Omega=0$, where $E$ is the diagonal matrix
\begin{equation}
E=\pm\left(\begin{array}{ccc}
1&0&0\\
0&1&0\\
0&0&\epsilon
\end{array}\right).\label{SUSYE}
\end{equation}

\subsection{Bosonic extension}
We now construct the SUSY GW and SUSY GC equations for the bosonic extension. From the relations (\ref{SUSYrelbij}), the coefficients $b^i_{\phantom{i}j}$ are given by
\be
\hspace{-1cm} b^1_{\phantom{1}1}=H,\qquad b^2_{\phantom{2}1}=2e^{-\phi}Q^+,\qquad b^1_{\phantom{1}2}=-2e^{-\phi}Q^-,\qquad b^2_{\phantom{2}2}=H,
\ee
up to an additional bosonic bodiless function $\zeta_1\neq0$ such that $\zeta_1f=0$ and where the $b^k_{\phantom{k}j}$ are the mixed coefficients of the second fundamental form. Therefore the SUSY GW equations (\ref{SUSYGW}) take the form
\be
\hspace{-2.5cm}\ba{cc}
D_+\Omega=A_+\Omega, & D_-\Omega=A_-\Omega, \\
A_+=\left(\ba{ccc}
\Gamma_{11}^{\phantom{11}1} & \Gamma_{11}^{\phantom{11}2} & Q^+f \\
-\Gamma_{12}^{\phantom{12}1} & -\Gamma_{12}^{\phantom{12}2} & -\frac{1}{2}e^{\phi}Hf \\
H & 2e^{-\phi}Q^+ & 0
\ea\right), & A_-=\left(\ba{ccc}
\Gamma_{12}^{\phantom{12}1} & \Gamma_{12}^{\phantom{12}2} & \frac{1}{2}e^\phi Hf \\
\Gamma_{22}^{\phantom{22}1} & \Gamma_{22}^{\phantom{22}2} & Q^-f \\
-2e^{-\phi}Q^- & H & 0
\ea\right),
\ea\label{BGW}
\ee
where the matrices $A_\pm$ are in the Bianchi form \cite{Bianchi}. The compatibility condition of the SUSY GW equations is
\be
D_+A_-+D_-A_+-\lbrace EA_+,EA_-\rbrace=0,\label{BZCC}
\ee
up to the addition of a non-zero matrix $P$ such that $P\Omega=0$, where
\begin{equation}
E=\pm\left(\begin{array}{ccc}
1&0&0\\
0&1&0\\
0&0&-1
\end{array}\right),\label{BE}
\end{equation}
Using the subblock notation, the matrices $A_\pm$ can also be written as
\begin{equation}
A_+=\left(\begin{array}{cc|c}
\Gamma_{11}^{\phantom{11}1} & \Gamma_{11}^{\phantom{11}2} & Q^+f \\
-\Gamma_{12}^{\phantom{12}1} & -\Gamma_{12}^{\phantom{12}2} & -\frac{1}{2}e^\phi Hf \\
\hline H & 2e^{-\phi}Q^+ & 0
\end{array}\right)=\left(\begin{array}{c|c}
A_f^+&I_{b_1}^+\\
\hline I_{b_2}^+&0
\end{array}\right),\label{BA+}
\end{equation}
\begin{equation}
A_-=\left(\begin{array}{cc|c}
\Gamma_{12}^{\phantom{12}1} & \Gamma_{12}^{\phantom{12}2} & \frac{1}{2}e^\phi Hf \\
\Gamma_{22}^{\phantom{22}1} & \Gamma_{22}^{\phantom{22}2} & Q^-f \\
\hline-2e^{-\phi}Q^- & H  & 0
\end{array}\right)=\left(\begin{array}{c|c}
A_f^-&I_{b_1}^-\\
\hline I_{b_2}^-&0
\end{array}\right),\label{BA-}
\end{equation}
where $A_f^+$ and $A_f^-$ are $2\times2$ matrices with fermionic entries, $I_{b_1}^+$ and $I_{b_1}^-$ are two-component column vectors with bosonic entries, and $I_{b_2}^+$ and $I_{b_2}^-$ are two-component row vectors with bosonic entries. Let us consider the moving frame $\Psi=(\psi_f,\psi_b)^T$ where $\psi_f$ is a two-component fermionic vector and $\psi_b$ is a bosonic scalar. From the GW equations for the moving frame $\Psi$, with the matrices given by (\ref{BA+}) and (\ref{BA-}), we obtain
\be
D_+\Psi=A_+\Psi,\qquad D_-\Psi=A_-\Psi.
\ee
The compatibility conditions for the $\psi_f$ and $\psi_b$ lead us to the four equations
\begin{equation}
\begin{array}{l}
D_+A^-_f+D_-A^+_f+I_{b_1}^-I_{b_2}^++I_{b_1}^+I_{b_2}^--\lbrace A_f^+,A_f^-\rbrace=0,\\
-A_f^-I_{b_1}^++D_+I_{b_1}^-+I_{b_1}^-\eta_f^+-A_f^+I_{b_1}^-+D_-I_{b_1}^++I_{b_1}^+\eta_f^-=0,\\
D_+I_{b_2}^-+I_{b_2}^-A_f^+-\eta_f^-I_{b_2}^++D_-I_{b_2}^++I_{b_2}^+A_f^--\eta_f^+I_{b_2}^-=0,\\
I_{b_2}^+I_{b_1}^-+D_-\eta_f^++I_{b_2}^-I_{b_1}^++D_+\eta_f^-=0.\label{BZCC2}
\end{array}
\end{equation}
The zero curvature condition (ZCC) corresponding to the equations (\ref{BZCC2}) is an equivalent matrix form of (\ref{BZCC}).

The ZCC (\ref{BZCC}) leads us to the bosonic SUSY GC equations which consist of the following six linearly independent equations for the matrix components
\be
\hspace{-2.5cm}\ba{ll}
(i) & D_-(\Gamma_{11}^{\phantom{11}1})+D_+(\Gamma_{22}^{\phantom{22}2})+D_+(\Gamma_{12}^{\phantom{12}1})-D_-(\Gamma_{12}^{\phantom{12}2})=0, \\
(ii) & D_-(\Gamma_{11}^{\phantom{11}1})-\Gamma_{11}^{\phantom{11}2}\Gamma_{22}^{\phantom{22}1} +D_+(\Gamma_{12}^{\phantom{12}1})+\Gamma_{12}^{\phantom{12}2}\Gamma_{12}^{\phantom{12}1}+\frac{1}{2}H^2e^\phi f-2Q^+Q^-e^{-\phi}f=0, \\
(iii) & Q^+\Gamma_{22}^{\phantom{22}2}-\Gamma_{11}^{\phantom{11}2}Q^-+D_-Q^+-Q^+D_-\phi+\frac{1}{2}e^\phi D_+H=0, \\
(iv) & Q^-\Gamma_{11}^{\phantom{11}1}-\Gamma_{22}^{\phantom{22}1}Q^++D_+Q^--Q^-D_+\phi-\frac{1}{2}e^\phi D_-H=0, \\
(v) & D_-(\Gamma_{11}^{\phantom{11}2})-\Gamma_{12}^{\phantom{12}1}\Gamma_{11}^{\phantom{11}2} -\Gamma_{11}^{\phantom{11}2}\Gamma_{22}^{\phantom{22}2}-\Gamma_{11}^{\phantom{11}1}\Gamma_{12}^{\phantom{12}2} +D_+(\Gamma_{12}^{\phantom{12}2})+2Q^+Hf=0, \\
(vi) & D_+(\Gamma_{22}^{\phantom{22}1})+\Gamma_{12}^{\phantom{12}2}\Gamma_{22}^{\phantom{22}1}-\Gamma_{22}^{\phantom{22}1} \Gamma_{11}^{\phantom{11}1}+\Gamma_{22}^{\phantom{22}2}\Gamma_{12}^{\phantom{12}1}-D_-(\Gamma_{12}^{\phantom{12}1})+2Q^-Hf=0.
\ea\label{BGC}
\ee
The Grassmann-valued PDEs (\ref{BGC}) involve eleven dependent functions of the independent variables $x_+$, $x_-$, $\theta^+$ and $\theta^-$  including the four bosonic functions $\phi$, $H$, $Q^\pm$ and the six fermionic functions $\Gamma_{ij}^{\phantom{ij}k}$ together with one dependent bodiless bosonic function $f$ of $x_+$ and $x_-$. It is interesting to note that the equation (\ref{BGC}.i) is the compatibility condition of the function $f$ given in equation (\ref{DDf}). Under the above assumptions we obtain the following result.

\begin{proposition}[Structural bosonic SUSY equations]\hfill\\
For any vector bosonic superfields $F(x_+,x_-,\theta^+,\theta^-)$ and $N(x_+,x_-,\theta^+,\theta^-)$ satisfying the normalization conditions (\ref{NormalizationF}), (\ref{gij}), (\ref{NormalizationN}) and (\ref{QHnorm}), the moving frame $\Omega=(D_+F,D_-F,N)^T$ on a smooth conformally parametrized surface immersed in the superspace $\mathbb{R}^{(2,1\vert2)}$ satisfies the bosonic SUSY GW equations (\ref{BGW}). The ZCC (\ref{BZCC}), which is the compatibility condition of the bosonic SUSY GW equations (\ref{BGW}) expressed in terms of the matrices $A_+$ and $A_-$, is equivalent to the bosonic SUSY GC equations (\ref{BGC}).
\end{proposition}

\subsection{Fermionic extension}
We now discuss a derivation of the SUSY GW and SUSY GC equations for the fermionic extension. Conditions on the Christoffel symbols of the second kind $\Gamma_{ij}^{\phantom{ij}k}$ can be obtained by taking derivatives of (\ref{NormalizationF}), i.e.
\be
\hspace{-2.5cm}0\hspace{-0.1cm}=\hspace{-0.1cm}D_i\langle D_jF,D_jF\rangle\hspace{-1mm}=\hspace{-1mm}\langle D_iD_jF,D_jF\rangle+\langle D_jF,D_iD_jF\rangle\hspace{-1mm}=\hspace{-1mm}2\langle D_iD_jF,D_jF\rangle\hspace{-1mm}=\hspace{-1mm}2\Gamma_{ji}^{\phantom{ji}k}g_{kj}.
\ee
Therefore we have
\be
\Gamma_{ji}^{\phantom{ji}k}=0,\qquad\mbox{for}\qquad j\neq i
\ee
and since the Christoffel symbols are antisymmetric under a permutation of indices $i$ and $j$ (see equation (\ref{antiChrist})), we get
\be
\Gamma_{ji}^{\phantom{ji}k}=0,\qquad \mbox{if}\qquad i\neq k\quad\mbox{or}\quad j\neq k.
\ee
Using the last result on equations (\ref{Df}) and (\ref{DDf}) we get
\be
\hspace{-1cm}D_kf=(\Gamma_{k(k)}^{\phantom{k(k)}(k)}-D_k\phi)f,\qquad \lbrace D_+,D_-\rbrace f=(D_+\Gamma_{22}^{\phantom{22}2}+D_-\Gamma_{11}^{\phantom{11}1})f=0.\label{FDDf}
\ee
Also the Christoffel symbols of the first kind are given by
\be\ba{llll}
\Gamma_{111}=0, & \Gamma_{112}=\frac{1}{2}e^\phi \Gamma_{11}^{\phantom{11}1}, & \Gamma_{121}=0, & \Gamma_{211}=0,\\
\Gamma_{122}=0, & \Gamma_{212}=0, & \Gamma_{221}=\frac{1}{2}e^\phi \Gamma_{22}^{\phantom{22}2}, & \Gamma_{222}=0,
\ea\ee
up to the addition of a fermionic function $\zeta_2\neq0$ which has the property $\zeta_2f=0$. The fermionic quantities $b^i_{\phantom{i}j}$ take the form
\be
\hspace{-1cm}b^1_{\phantom{1}1}=H,\qquad b^2_{\phantom{2}1}=-2e^{-\phi}Q^+,\qquad b^1_{\phantom{1}2}=-2e^{-\phi}Q^-,\qquad b^2_{\phantom{2}2}=-H,
\ee
up to the addition of a fermionic function $\zeta_3\neq0$ which has the property $\zeta_3f=0$.
Therefore, the fermionic-valued matrices $A_\pm$ in the SUSY GW equations (\ref{SUSYGW}) take the form
\be
\ba{c}
D_+\left(\ba{c}D_+F\\D_-F\\N\ea\right)=\left(\ba{ccc}
\Gamma_{11}^{\phantom{11}1} & 0 & Q^+f \\
0 & 0 & -\frac{1}{2}e^\phi Hf \\
H & -2e^{-\phi}Q^+ & 0
\ea\right)\left(\ba{c}D_+F\\D_-F\\N\ea\right),\\
\\
D_-\left(\ba{c}D_+F\\D_-F\\N\ea\right)=\left(\ba{ccc}
0 & 0 & \frac{1}{2}e^\phi Hf \\
0 & \Gamma_{22}^{\phantom{22}2} & Q^-f \\
-2e^{-\phi}Q^- & -H & 0
\ea\right)\left(\ba{c}D_+F\\D_-F\\N\ea\right).
\ea\label{FGW}
\ee
The compatibility condition of the SUSY GW equations (\ref{FGW}) are given by
\be
D_+A_-+D_-A_+-\lbrace A_+,A_-\rbrace=0,\label{FZCC}
\ee
up to the addition of a non-zero matrix $P$ such that $P\Omega=0$. In component form these equations are
\be\ba{rl}
(i) & D_-(\Gamma_{11}^{\phantom{11}1})+2e^{-\phi}Q^+Q^-f=0,\\
(ii) & \left[D_-Q^++\frac{1}{2}e^\phi D_+H+Q^+(D_-\phi-\Gamma_{22}^{\phantom{22}2})\right]f=0,\\
(iii) & D_+(\Gamma_{22}^{\phantom{22}2})+2e^{-\phi}Q^-Q^+f=0,\\
(iv) & \left[D_+Q^--\frac{1}{2}e^\phi D_-H+Q^-(D_+\phi-\Gamma_{11}^{\phantom{11}1})\right]f=0,\\
(v) & D_+Q^--\frac{1}{2}e^\phi D_-H+Q^-(D_+\phi-\Gamma_{11}^{\phantom{11}1})=0,\\
(vi) & D_-Q^++\frac{1}{2}e^\phi D_+H+Q^+(D_-\phi-\Gamma_{22}^{\phantom{22}2})=0.
\ea\label{FpreGC}\ee
If the equations (\ref{FpreGC}.vi) and (\ref{FpreGC}.v) hold, then equations (\ref{FpreGC}.ii) and (\ref{FpreGC}.iv) are identically satisfied respectively. By adding (\ref{FpreGC}.i) and (\ref{FpreGC}.iii) we obtain
\be
D_-(\Gamma_{11}^{\phantom{11}1})+D_+(\Gamma_{22}^{\phantom{22}2})=0,
\ee
which is the compatibility condition for $f$, corresponding to (\ref{FDDf}). Therefore the SUSY GC equations are reduced to the four linearly independent equations
\be\ba{rl}
(i) & D_+(\Gamma_{22}^{\phantom{22}2})+D_-(\Gamma_{11}^{\phantom{11}1})=0,\\
(ii) & D_-(\Gamma_{11}^{\phantom{11}1})+2e^{-\phi}Q^+Q^-f=0,\\
(iii) & D_+Q^--\frac{1}{2}e^\phi D_-H+Q^-(D_+\phi-\Gamma_{11}^{\phantom{11}1})=0,\\
(iv) & D_-Q^++\frac{1}{2}e^\phi D_+H+Q^+(D_-\phi-\Gamma_{22}^{\phantom{22}2})=0.
\ea\label{FGC}\ee
The Grassmann-valued PDEs (\ref{FGC}) involve six dependent functions of the independent variables $x_+$, $x_-$, $\theta^+$ and $\theta^-$ including one bosonic function $\phi$ and the five fermionic functions $\Gamma_{11}^{\phantom{11}1}$, $\Gamma_{22}^{\phantom{22}2}$, $H$ and $Q^\pm$ together with one bosonic function $f$ of $x_+$ and $x_-$. 

\begin{proposition}[Structural fermionic SUSY equations]\hfill\\
For any vector fermionic superfield $F(x_+,x_-,\theta^+,\theta^-)$ and bosonic superfield $N(x_+,x_-,\theta^+,\theta^-)$ satisfying the normalization conditions (\ref{NormalizationF}), (\ref{gij}), (\ref{NormalizationN}) and (\ref{QHnorm}), the bosonic moving frame $\Omega=(D_+F,D_-F,N)^T$ on a smooth conformally parametrized surface immersed in the superspace $\mathbb{R}^{(1,1\vert3)}$ satisfies the fermionic SUSY GW equations (\ref{FGW}). The ZCC (\ref{FZCC}), which is the compatibility condition of the fermionic SUSY GW equations (\ref{FGW}) expressed in terms of the matrices $A_+$ and $A_-$, is equivalent to the fermionic SUSY GC equations (\ref{FGC}).
\end{proposition}

If we consider the case where $f$ is a bosonic constant in the fermionic extension, then from equations (\ref{FDDf}) we have that
\be
\Gamma_{11}^{\phantom{11}1}=D_+\phi,\qquad \Gamma_{22}^{\phantom{22}2}=D_-\phi,
\ee
up to the addition of a function $\zeta_4\neq0$ with the property $\zeta_4 f=0$. Also the compatibility condition on $f$ is then identically satisfied. The SUSY GC equations become
\be\ba{rl}
(i) & D_-D_+\phi+2e^{-\phi}Q^+Q^-f=0,\\
(ii) & D_+Q^--\frac{1}{2}e^\phi D_-H=0,\\
(iii) & D_-Q^++\frac{1}{2}e^\phi D_+H=0,
\ea\label{FGC0}\ee
which resemble the classical GC equations (\ref{ClaGC}) taking into account that the $H^2$ term vanishes. The equations (\ref{FGC0}) contain terms whose signs differ from those of the classical equations. We get an underdetermined system of three PDEs for four dependent variables $H$, $Q^\pm$ and $\phi$. One should note that for this special case the SUSY GW equations
\be
\ba{c}
D_+\left(\ba{c}D_+F\\D_-F\\N\ea\right)=\left(\ba{ccc}
D_+\phi & 0 & Q^+f \\
0 & 0 & -\frac{1}{2}e^\phi Hf \\
H & -2e^{-\phi}Q^+ & 0
\ea\right)\left(\ba{c}D_+F\\D_-F\\N\ea\right),\\
\\
D_-\left(\ba{c}D_+F\\D_-F\\N\ea\right)=\left(\ba{ccc}
0 & 0 & \frac{1}{2}e^\phi Hf \\
0 & D_-\phi & Q^-f \\
-2e^{-\phi}Q^- & -H & 0
\ea\right)\left(\ba{c}D_+F\\D_-F\\N\ea\right),
\ea\label{FGW0}
\ee
are also similar to the classical GW equations up to some sign differences and the multiplication of some elements by the function $f$.

\section{Geometric aspects of conformally parametrized SUSY surfaces}\setcounter{equation}{0}
In this section, we discuss certain aspects of Grassmann variables in conjunction with differential geometry and SUSY analysis. Let us define the differential fermionic operators
\be
d_\pm=d\theta^\pm+idx_\pm\partial_{\theta^\pm},
\ee
where $d_+$ and $d_-$ are the infinitesimal displacements in the direction of $D_+$ and $D_-$, respectively. These operators are anticommuting, i.e. $\lbrace d_+,d_-\rbrace=0$. In order to compute the first and second fundamental forms, we have assumed that $(d\theta^j\_\hspace{-0.15cm}\shortmid\partial_{\theta^i})=0$ for $i, j=1,2.$ For SUSY conformally parametrized surfaces, the first fundamental form is given by
\be
\hspace{-1.5cm}\ba{l}
I=\left\langle\left(\ba{cc}d_+ & d_-\ea\right),\left(\ba{cc}d_+ & d_-\ea\right)\left(\ba{cc}
\langle D_+F,D_+F\rangle & \langle D_+F,D_-F\rangle\\
-\langle D_+F,D_-F\rangle & \langle D_-F,D_-F\rangle
\ea\right)\right\rangle\\
\phantom{I}=\left\langle\left(\ba{cc}d_+ & d_-\ea\right),\left(\ba{cc}d_+ & d_-\ea\right)\left(\ba{cc}
g_{11}f & g_{12}f\\
-g_{12}f & g_{22}f
\ea\right)\right\rangle\\
\phantom{I}=\left\langle\left(\ba{cc}d_+ & d_-\ea\right),\left(\ba{cc}d_+ & d_-\ea\right)Rf\right\rangle\\
\phantom{I}=f\left(d_+^2g_{11}+2d_+d_-g_{12}+d_-^2g_{22}\right)=fd_+d_-e^\phi,\label{SUSYI}
\ea
\ee
where the $2\times2$ bosonic-valued matrix $R$ is given by
\be
R=\left(\ba{cc}
g_{11} & g_{12}\\
-g_{12} & g_{22}
\ea\right)=\frac{1}{2}e^\phi\left(\ba{cc}
0 & 1\\
-1 & 0
\ea\right).
\ee
The discriminant $g$ is defined to be
\be
g=g_{11}g_{22}+(g_{12})^2=\frac{1}{4}e^{2\phi}.\label{SUSYg}
\ee
Hence the covariant metric is given by
\be
g_{ij}g^{jk}=\delta_i^k,\qquad \left(\ba{cc}
g_{11} & g_{12} \\
g_{21} & g_{22}
\ea\right)\left(\ba{cc}
g^{11} & g^{12} \\
g^{21} & g^{22}
\ea\right)=\left(\ba{cc}
1 & 0 \\
0 & 1
\ea\right),
\ee
such that
\be
g^{11}=g^{22}=0,\qquad g^{12}=\epsilon g^{21}=2e^{-\phi},
\ee
where $\epsilon=-1$ in the bosonic case and $\epsilon=1$ in the fermionic case. The second fundamental form is
\be
\hspace{-1.5cm}\ba{l}
I\hspace{-0.1cm}I=\left\langle\left(\ba{cc}d_+ & d_-\ea\right),\left(\ba{cc}d_+ & d_-\ea\right)\left(\ba{cc}
\langle D_+^2F,N\rangle & \langle D_-D_+F,N\rangle\\
-\langle D_-D_+F,N\rangle & \langle D_-^2F,N\rangle
\ea\right)\right\rangle\\
\phantom{I}=\left\langle\left(\ba{cc}d_+ & d_-\ea\right),\left(\ba{cc}d_+ & d_-\ea\right)\left(\ba{cc}
b_{11}f & b_{12}f\\
-b_{12}f & b_{22}f
\ea\right)\right\rangle\\
\phantom{I}=\left\langle\left(\ba{cc}d_+ & d_-\ea\right),\left(\ba{cc}d_+ & d_-\ea\right)Sf\right\rangle\\
\phantom{I}=f\left(d_+^2b_{11}+2d_+d_-b_{12}+d_-^2b_{22}\right)=f\left(d_+^2Q^++d_+d_-(e^\phi H)+d_-^2Q^-\right),
\ea\label{SUSYII}
\ee
where the matrix $S$ is given by
\be
S=\left(\ba{cc}
b_{11} & b_{12}\\
-b_{12} & b_{22}
\ea\right)=\left(\ba{cc}
Q^+ & \frac{1}{2}e^\phi H\\
-\frac{1}{2}e^\phi H & Q^-
\ea\right).
\ee
The discriminant $b$ is defined to be
\be
b=b_{11}b_{22}+(b_{12})^2=Q^+Q^-+\frac{1}{4}e^{2\phi}H^2.\label{SUSYb}
\ee 
One should note that the term in $H^2$ vanishes in the fermionic case since $H$ is a fermionic-valued function.
Making use of (\ref{SUSYg}) and (\ref{SUSYb}), the Gaussian and mean curvatures are defined as
\be
\hspace{-2.5cm}\mathcal{K}=\det(SR^{-1})=\frac{b_{11}b_{22}+(b_{12})^2}{g_{11}g_{22}+(g_{12})^2}=4e^{-2\phi}Q^+Q^-+H^2,\qquad H=\frac{1}{2}\mbox{tr}(SR^{-1}).\label{SUSYKH}
\ee
In the fermionic case, $\mathcal{K}$ is a bosonic bodiless function and the term $H^2$ vanishes. 

\paragraph{}Under the above assumptions on the SUSY versions of the GC equations (\ref{BGC}) or (\ref{FGC}) we can provide a SUSY analogue of the Bonnet theorem.
\begin{proposition}[SUSY extension of the Bonnet theorem]\hfill\\
Given a SUSY conformal metric, $M=fd_+d_-e^\phi$, of a smooth conformally parametrized surface $\mathcal{S}$, the Hopf differentials $d_\pm^2Q^\pm $ and a mean curvature function $H$ defined on a Riemann surface $\mathcal{R}$ satisfying the GC equations ((\ref{BGC}) for the bosonic case, (\ref{FGC}) for the fermionic case), there exists a vector-valued immersion function, 
\be
\ba{l}
F^b=(F_1^b,F_2^b,F_3^b):\tilde{\mathcal{R}}\rightarrow\mathbb{R}^{(2,1\vert2)},\\
F^f=(F_1^f,F_2^f,F_3^f):\tilde{\mathcal{R}}\rightarrow\mathbb{R}^{(1,1\vert3)},
\ea
\ee
(bosonic or fermionic, respectively) with the fundamental forms
\begin{equation}
I= fd_+d_-e^\phi,\qquad I\hspace{-0.1cm}I=f(d_+^2Q^++ d_+d_-(He^\phi)+d_-^2Q^-),
\end{equation}
where $\tilde{\mathcal{R}}$ is the universal covering of the Riemann surface $\mathcal{R}$ and $\mathbb{R}^{(n_b\vert n_f)}$ is the superspace. The immersion function $F$ is unique up to affine transformations in the superspace.
\end{proposition}

\paragraph{}The proof of this proposition is analogous to that given in \cite{Bonnet}. Note that it is straightforward to construct surfaces in the superspace related to integrable equations. However, it is non-trivial to identify those surfaces which have an invariant geometrical characterization. A list of such surfaces is known in the classical case \cite{Bob} but, to our knowledge, an identification of such surfaces is an open problem in the case of surfaces immersed in the superspace.

\section{Symmetries of the SUSY GC equations}\setcounter{equation}{0}
By a symmetry supergroup $G$ of a SUSY system, we mean a local supergroup of transformations acting on the Cartesian product $\mathcal{X}\times\mathcal{U}$ of supermanifolds, where $\mathcal{X}$ is the space of four independent variables $(x_+,x_-,\theta^+,\theta^-)$ and $\mathcal{U}$ is the space of dependent superfields. For the bosonic case, $\mathcal{U}$ is the space of eleven dependent superfields $\mathcal{U}=(\phi,H,Q^+,Q^-,R^+,R^-,S^+,S^-,T^+,T^-,f)$, where we have used the abbreviated notation for the Christoffel symbols of the second kind
\be
\hspace{-2cm}R^+=\Gamma_{11}^{\phantom{11}1},\quad R^-=\Gamma_{11}^{\phantom{11}2},\quad S^+=\Gamma_{12}^{\phantom{12}1},\quad S^-=\Gamma_{12}^{\phantom{12}2},\quad T^+=\Gamma_{22}^{\phantom{22}1},\quad T^-=\Gamma_{22}^{\phantom{22}2}.\label{NotationGamma}
\ee
For the fermionic case, $\mathcal{U}$ is the space of seven dependent superfields $\mathcal{U}=(\phi,H,Q^+,Q^-,R^+,T^-,f)$, where we have used the notation (\ref{NotationGamma}) for the non-zero Christoffel symbols of the second kind $\Gamma_{ij}^{\phantom{ij}k}$. Solutions of the SUSY GC equations, (\ref{BGC}) for the bosonic case or (\ref{FGC}) for the fermionic case, are mapped to solutions of equations (\ref{BGC}) or (\ref{FGC}), respectively by the action of the supergroup $G$ on the functions in $\mathcal{U}$. When we perform the symmetry reductions, we need to take into consideration the fact that the bosonic function $f$ introduced in (\ref{NormalizationF}) depends only on $x_+$ and $x_-$ or is constant. If $G$ is a Lie supergroup as described in \cite{Kac} and \cite{Winternitz}, it can be associated with its Lie superalgebra whose elements are infinitesimal symmetries of the SUSY GC equations. We have made use of the theory described in the book by Olver \cite{Olver} in order to determine superalgebras of infinitesimal symmetries for both the bosonic and fermionic SUSY GC equations.

The bosonic SUSY GC equations (\ref{BGC}) are invariant under the Lie superalgebra $\mathfrak{g}$ generated by the following eight infinitesimal vector fields
\be
\hspace{-2.5cm}\ba{l}
C_0= H\partial_H+Q^+\partial_{Q^+}+Q^-\partial_{Q^-}-2f\partial_f,\\
K_0= -H\partial_H+Q^+\partial_{Q^+}+Q^-\partial_{Q^-}+2\partial_\phi,\\
K_1^b=-2x_+\partial_{x_+}-\theta^+\partial_{\theta^+}+R^+\partial_{R^+}+2R^-\partial_{R^-}+S^-\partial_{S^-}-T^+\partial_{T^+}+2Q^+\partial_{Q^+}+\partial_\phi,\\
K_2^b=-2x_-\partial_{x_-}-\theta^-\partial_{\theta^-}-R^-\partial_{R^-}+S^+\partial_{S^+}+2T^+\partial_{T^+}+ T^-\partial_{T^-}+2Q^-\partial_{Q^-}+\partial_\phi,\\
P_+=\partial_{x_+},\qquad\qquad\qquad\qquad P_-=\partial_{x_-},\\
J_+=\partial_{\theta^+}+i\theta^+\partial_{x_+},\qquad\qquad J_-=\partial_{\theta^-}+i\theta^-\partial_{x_-}.
\ea\label{BsymGC}
\ee
The generators $P_+$ and $P_-$ represent translations in the bosonic variables $x_+$ and $x_-$ while $K_1^b$, $K_2^b$, $K_0$ and $C_0$ generate dilations on both even and odd variables. In addition, we recover the SUSY operators $J_+$ and $J_-$ which were identified previously in equation (\ref{J}). The commutation (anticommutation in the case of two fermionic operators) relations of the superalgebra $\mathfrak{g}$ of the SUSY GC equations (\ref{BGC}) are given in table~1 for the case $D_\pm f\neq0$. 
\begin{table}
\centering
\caption{Commutation table for the Lie superalgebra $\mathfrak{g}$ spanned by\\ the vector fields (\ref{BsymGC}). In the case of two fermionic generators $J_+$\\ and/or $J_-$ we have anticommutation rather than commutation.}
\begin{tabular}{c|c|c|c|c|c|c|c|c|}
 &$K_1^b$&$P_+$&$J_+$&$K_2^b$&$P_-$&$J_-$&$K_0$&$C_0$\\
\hline$K_1^b$&$0$ &$2P_+$ &$J_+$ &$0$ &$0$ &$0$ &$0$ &$0$ \\
\hline$P_+$&$-2P_+$ &$0$ &$0$ &$0$ &$0$ &$0$ &$0$ &$0$ \\
\hline$J_+$&$-J_+$ &$0$ &$2iP_+$ &$0$ &$0$ &$0$ &$0$ &$0$ \\
\hline$K_2^b$&$0$ &$0$ &$0$ &$0$ &$2P_-$ &$J_-$ &$0$ &$0$ \\
\hline$P_-$&$0$ &$0$ &$0$ &$-2P_-$ &$0$ &$0$ &$0$ &$0$ \\
\hline$J_-$&$0$ &$0$ &$0$ &$-J_-$ &$0$  &$2iP_-$ &$0$ &$0$ \\
\hline$K_0$&$0$ &$0$ &$0$ &$0$ &$0$ &$0$ &$0$ &$0$ \\
\hline$C_0$&$0$ &$0$ &$0$ &$0$ &$0$ &$0$ &$0$ &$0$ \\
\hline
\end{tabular}
\end{table}
The Lie superalgebra $\mathfrak{g}$ can be decomposed into the following combination of direct and semi-direct sums
\begin{equation}
\mathfrak{g}=\lbrace\lbrace K_1^b\rbrace\sdir\lbrace P_+,J_+\rbrace\rbrace \oplus\lbrace\lbrace K_2^b\rbrace\sdir\lbrace P_-,J_-\rbrace\rbrace\oplus\lbrace K_0\rbrace\oplus\lbrace C_0\rbrace.\label{Bdec}
\end{equation}
In equation (\ref{Bdec}) the brace bracket $\lbrace\cdot,\cdot\rbrace$ denotes the set of all elements within. It should be noted that $K_0$ and $C_0$ constitute the center of the Lie superalgebra $\mathfrak{g}$.

The fermionic SUSY GC equations (\ref{FGC}) are invariant under the following six bosonic symmetry generators
\renewcommand{\theequation}{\thesection.\arabic{equation}a}
\be
\ba{l}
P_+=\partial_{x_+},\qquad P_-=\partial_{x_-},\\
C_0=H\partial_H+Q^+\partial_{Q^+}+Q^-\partial_{Q^-}-2f\partial_f,\\
K_0=-H\partial_H+Q^+\partial_{Q^+}+Q^-\partial_{Q^-}+2\partial_\phi,\\
K_1^f=-2x_+\partial_{x_+}-\theta^+\partial_{\theta^+}+2Q^+\partial_{Q^+}+R^+\partial_{R^+}+\partial_\phi,\\
K_2^f=-2x_-\partial_{x_-}-\theta^-\partial_{\theta^-}+2Q^-\partial_{Q^-}+T^-\partial_{T^-}+\partial_\phi,
\ea\label{FsymGCbos}
\ee
together with the three fermionic generators\setcounter{equation}{3}\renewcommand{\theequation}{\thesection.\arabic{equation}b}
\be
J_+=\partial_{\theta^+}+i\theta^+\partial_{x_+},\qquad J_-=\partial_{\theta^-}+i\theta^-\partial_{x_-},\qquad W=\partial_H.\label{FsymGCfer}
\ee
\renewcommand{\theequation}{\thesection.\arabic{equation}}
The symmetry generators $W$ and $P_\pm$ represent a fermionic translation of $H$ and bosonic translations in the $x_\pm$ direction respectively, $J_\pm$ represent the SUSY transformations and $C_0$, $K_0$, $K_1^f$ and $K_2^f$ represent dilations. The commutation table (anticommutation for two fermionic symmetries) for the generators of the superalgebra $\mathfrak{h}$ of equations (\ref{FGC}) is given in table~2.
\begin{table}
\centering
\caption{Commutation table for the Lie superalgebra $\mathfrak{h}$ spanned by\\ the vector fields (6.4). In the case of two fermionic generators $J_+$\\ and/or $J_-$ and/or $W$ we have anticommutation rather than commutation.}
\begin{tabular}{c|c|c|c|c|c|c|c|c|c|}
 &$K_1^f$&$P_+$&$J_+$&$K_2^f$&$P_-$&$J_-$&$K_0$&$C_0$&$W$\\
\hline$K_1^f$&$0$ &$2P_+$ &$J_+$ &$0$ &$0$ &$0$ &$0$&$0$&$0$ \\
\hline$P_+$&$-2P_+$ &$0$ &$0$ &$0$ &$0$ &$0$ &$0$&$0$&$0$ \\
\hline$J_+$&$-J_+$ &$0$ &$2iP_+$ &$0$ &$0$ &$0$ &$0$&$0$&$0$ \\
\hline$K_2^f$&$0$ &$0$ &$0$ &$0$ &$2P_-$ &$J_-$ &$0$&$0$&$0$ \\
\hline$P_-$&$0$ &$0$ &$0$ &$-2P_-$ &$0$ &$0$ &$0$&$0$&$0$ \\
\hline$J_-$&$0$ &$0$ &$0$ &$-J_-$ &$0$  &$2iP_-$ &$0$&$0$&$0$ \\
\hline$K_0$&$0$ &$0$ &$0$ &$0$ &$0$ &$0$ &$0$&$0$&$W$ \\
\hline$C_0$&$0$ &$0$ &$0$ &$0$ &$0$ &$0$ &$0$&$0$&$-W$ \\
\hline$W$&$0$ &$0$ &$0$ &$0$ &$0$ &$0$ &$-W$&$W$ &$0$ \\
\hline
\end{tabular}
\end{table}
The decomposition of the superalgebra (6.4) is given by
\be
\hspace{-1cm}\mathfrak{h}=\lbrace\lbrace K_1^f\rbrace\sdir\lbrace P_+,J_+\rbrace\rbrace \oplus\lbrace\lbrace K_2^f\rbrace\sdir\lbrace P_-,J_-\rbrace\rbrace\oplus\lbrace\lbrace K_0,C_0\rbrace\sdir\lbrace W\rbrace\rbrace.\label{Fdec}
\ee

However, if we consider the case where $D_\pm f=0$, the equations (\ref{FGC0}) are invariant under the five bosonic generators\renewcommand{\theequation}{\thesection.\arabic{equation}a}
\be
\ba{l}
P_+=\partial_{x_+},\qquad P_-=\partial_{x_-},\\
K_0=-H\partial_H+Q^+\partial_{Q^+}+Q^-\partial_{Q^-}+2\partial_\phi,\\
\hat{K}_1^f=-2x_+\partial_{x_+}-\theta^+\partial_{\theta^+}+2Q^+\partial_{Q^+}+\partial_\phi,\\
\hat{K}_2^f=-2x_-\partial_{x_-}-\theta^-\partial_{\theta^-}+2Q^-\partial_{Q^-}+\partial_\phi,
\ea\label{FsymGC0}
\ee
and the three fermionic generators given by\setcounter{equation}{5}\renewcommand{\theequation}{\thesection.\arabic{equation}b}
\be
J_+=\partial_{\theta^+}+i\theta^+\partial_{x_+},\qquad J_-=\partial_{\theta^-}+i\theta^-\partial_{x_-},\qquad W=\partial_H.
\ee
The commutation table for the generators of the superalgebra of equations  (\ref{FGC0}) is given in table 3.\renewcommand{\theequation}{\thesection.\arabic{equation}}
\begin{table}
\centering
\caption{Commutation table for the Lie superalgebra $\hat{\mathfrak{h}}$ spanned by\\ the vector fields (6.6). In the case of two fermionic generators $J_+$\\ and/or $J_-$ and/or $W$ we have anticommutation rather than commutation.}
\begin{tabular}{c|c|c|c|c|c|c|c|c|}
 &$\hat{K}_1^f$&$P_+$&$J_+$&$\hat{K}_2^f$&$P_-$&$J_-$&$K_0$&$W$\\
\hline$\hat{K}_1^f$&$0$ &$2P_+$ &$J_+$ &$0$ &$0$ &$0$ &$0$&$0$ \\
\hline$P_+$&$-2P_+$ &$0$ &$0$ &$0$ &$0$ &$0$ &$0$&$0$ \\
\hline$J_+$&$-J_+$ &$0$ &$2iP_+$ &$0$ &$0$ &$0$ &$0$&$0$ \\
\hline$\hat{K}_2^f$&$0$ &$0$ &$0$ &$0$ &$2P_-$ &$J_-$ &$0$&$0$ \\
\hline$P_-$&$0$ &$0$ &$0$ &$-2P_-$ &$0$ &$0$ &$0$&$0$ \\
\hline$J_-$&$0$ &$0$ &$0$ &$-J_-$ &$0$  &$2iP_-$ &$0$&$0$ \\
\hline$K_0$&$0$ &$0$ &$0$ &$0$ &$0$ &$0$ &$0$&$W$ \\
\hline$W$&$0$ &$0$ &$0$ &$0$ &$0$ &$0$ &$-W$& $0$ \\
\hline
\end{tabular}
\end{table}
The Lie superalgebra $\hat{\mathfrak{h}}$, generated by (6.6), can be decomposed into the following combination of direct and semi-direct sums
\begin{equation}
\hat{\mathfrak{h}}=\lbrace\lbrace \hat{K}_1^f\rbrace\sdir\lbrace P_+,J_+\rbrace\rbrace \oplus\lbrace\lbrace \hat{K}_2^f\rbrace\sdir\lbrace P_-,J_-\rbrace\rbrace\oplus\lbrace\lbrace K_0\rbrace\sdir\lbrace W\rbrace\rbrace.\label{Fdec0}
\end{equation}

In both the bosonic and fermionic cases, the one-dimensional subalgebras of the respective superalgebra can be classified into conjugacy classes, as we proceed to do in the next section.

\section{One-dimensional subalgebras of the symmetry superalgebras of the SUSY GC equations}\setcounter{equation}{0}
In this section, we perform a classification of the one-dimensional subalgebras of the Lie superalgebras of infinitesimal transformations $\mathfrak{g}$ and $\mathfrak{h}$ into conjugacy classes under the action of their respective supergroups, $\exp(\mathfrak{g})$ generated by (\ref{BsymGC}), and $\exp(\mathfrak{h})$ generated by (6.4). The significance of such a classification resides in the fact that conjugate subgroups necessarily lead to invariant solutions which are equivalent in the sense that they can be transformed from one to the other by a suitable symmetry. Therefore, it is not necessary to compute reductions with respect to algebras which are conjugate to each other.

When constructing a list of representative one-dimensional subalgebras, it would be inconsistent to consider the $\mathbb{R}$ or $\mathbb{C}$ span of the generators (\ref{BsymGC}) or (6.4) because we multiply the odd generators $J_+$, $J_-$ and (in the fermionic case) $W$ by the odd parameters $\underline{\mu}$, $\underline{\eta}$ and $\underline{\zeta}$ respectively in the classifications listed in the Appendix. Therefore, one is naturally led to consider a superalgebra which is a supermanifold in the sense presented in section 3. This means that $\mathfrak{g}$ and $\mathfrak{h}$ contain any sums of even combinations of the bosonic generators (i.e. multiplied by even parameters including real or complex numbers) and odd combinations of fermionic generators (i.e. multiplied by odd parameters in $\Lambda_{odd}$). At the same time $\mathfrak{g}$ and $\mathfrak{h}$ are $\Lambda_{even}$ Lie modules. This fact can lead to the following complication. For a given $X$ in $\mathfrak{g}$ or $\mathfrak{h}$, the subalgebras $\mathfrak{X}$ and $\mathfrak{X}'$ spanned by $X$ and $X'=aX$ with $a\in\Lambda_{even}\backslash\mathbb{C}$ are not isomorphic in general, i.e. $\mathfrak{X}'\subset\mathfrak{X}$.

It should be noted that subalgebras obtained by multiplying other subalgebras by bodiless elements of $\Lambda_{even}$ do not provide us with anything new for the purpose of symmetry reduction. It is not particularly useful to consider a subalgebra of the form e.g. $\lbrace P_++\underline{\eta}_1\underline{\eta}_2P_-\rbrace$, since there is no limit to the number of odd parameters $\underline{\eta}_k$ that can be used to construct even coefficients. While such subalgebras may allow for more freedom in the choice of invariants, we then encounter the problem of non-standard invariants \cite{GHS09},\cite{GHS11}. Such non-standard invariants, which do not lead to standard reductions or invariant solutions, are found for several other SUSY hydrodynamic-type systems, e.g. in \cite{GH11},\cite{GH13}.

In what follows, we will assume throughout the computation of the non-isomorphic one-dimensional subalgebras that the non-zero bosonic parameters are invertible (i.e. behave essentially like ordinary real or complex numbers.) In order to classify the Lie superalgebras (\ref{BsymGC}) and (6.4) under the action of their respective supergroups, we make use of the techniques for classifying direct and semi-direct sums of algebras described in \cite{Winternitz} and generalize them to superalgebras involving both even and odd generators. In the case of direct sums, we use the Goursat twist method generalized to the case of a superalgebra. 

In the bosonic case, the superalgebra (\ref{Bdec}) contains two isomorphic copies of the 3-dimensional algebra $\mathfrak{g}^{(1)}=\lbrace\lbrace K_1^b\rbrace \sdir\lbrace P_+,J_+\rbrace\rbrace$ (the other copy being $\mathfrak{g}^{(2)}=\lbrace\lbrace K_2^b\rbrace \sdir\lbrace P_-,J_-\rbrace\rbrace$) together~with the one-dimensional algebras $\lbrace K_0\rbrace$ and $\lbrace C_0\rbrace$, which constitute the center of the Lie superalgebra $\mathfrak{g}$. This fact allows us to adapt the classification for 3-dimensional algebras as described in \cite{Patera}. So we begin our classification by considering the twisted one-dimensional subalgebras of $\mathfrak{g}^{(1)}\oplus\mathfrak{g}^{(2)}$. Under the action of a one-parameter group generated by the vector field
\begin{equation}
X=\alpha K_1^b+\beta P_++\underline{\eta}J_++\delta K_2^b+\lambda P_-+\underline{\rho}J_-,\label{7.3}
\end{equation}
where $\alpha,\beta,\delta,\lambda\in\Lambda_{even}$ and $\underline{\eta},\underline{\rho}\in\Lambda_{odd}$, the one-dimensional subalgebra
\begin{equation*}
Y=P_++aP_-,\qquad a\in\Lambda_{even}
\end{equation*}
transforms under the Baker-Campbell-Hausdorff formula
\begin{equation}
\hspace{-2cm}Y\rightarrow \mbox{Ad}_{\exp(X)}Y=Y+[X,Y]+\frac{1}{2!}[X,[X,Y]]+\frac{1}{3!}[X,[X,[X,Y]]]+...\label{BCH}
\end{equation}
to $e^{-2\alpha}P_++e^{-2\delta}aP_-$. Hence we get that $\lbrace P_++aP_-\rbrace$ is isomorphic to $\lbrace P_++e^{2\alpha-2\delta}aP_-\rbrace$. By a suitable choice of $\alpha$ and $\delta$, the factor $e^{2\alpha-2\delta}a$ can be rescaled to either $1$ or $-1$. Hence, we obtain a twisted subalgebra $\mathfrak{g}_{14}=\lbrace P_++\epsilon P_-,\epsilon=\pm1\rbrace$ given in table~4 in the Appendix.

As another example, consider a twisted subalgebra of the form $\lbrace P_++aK_2^b,~a\neq0\rbrace$, where $a\in\Lambda_{even}$. Through the Baker-Campbell-Hausdorff formula (\ref{BCH}), the vector field $Y=K_2^b+aP_+$ transforms (through the vector field $X$ given in (\ref{7.3})) to
\begin{equation}
e^XYe^{-X}=K_2^b+e^{-2\alpha}aP_+-\frac{\lambda}{\delta}(e^{-2\delta}-1)P_--\frac{1}{\delta}(e^{-\delta}-1)\underline{\rho}J_-.\label{BCH2}
\end{equation}
Through a suitable choice of $\lambda$ and $\underline{\rho}$, the last two terms of (\ref{BCH2}) can be eliminated, so we obtain the twisted subalgebra $\mathfrak{g}_{13}=\lbrace K_2^b+\epsilon P_+,\epsilon=\pm1\rbrace$. Continuing the classification in an analogous way, we obtain the list of one-dimensional subalgebras given in table~4 in the Appendix. 

In the fermionic case, the superalgebra (\ref{Fdec}) contains two isomorphic copies of the 3-dimensional algebra $\mathfrak{h}^{(1)}=\lbrace\lbrace K_1^f\rbrace\sdir\lbrace P_+,J_+\rbrace\rbrace$, the other copy being $\mathfrak{h}^{(2)}=\lbrace\lbrace K_2^f\rbrace\sdir\lbrace P_-,J_-\rbrace\rbrace$ together with the three-dimensional algebra $\lbrace\lbrace K_0,C_0\rbrace\sdir\lbrace W\rbrace\rbrace$. Therefore, we begin our classification by considering the twisted one-dimensional subalgebras of $\mathfrak{h}^{(1)}\oplus\mathfrak{h}^{(2)}$. Under the action of a one-parameter group generated by the vector field
\be
X=\alpha K_1^f+\beta P_++\underline{\eta}J_++\delta K_2^f+\lambda P_-+\underline{\rho}J_-,
\ee
where $\alpha,\beta,\delta,\lambda\in\Lambda_{even}$ and $\underline{\eta},\underline{\rho}\in\Lambda_{odd}$, the one-dimensional subalgebra $Y=P_++aP_-,a\in\Lambda_{even}$ transforms under the Baker-Campbell-Hausdorff formula (\ref{BCH}) to  $e^{-2\alpha}P_++e^{-2\delta}aP_-$. By an appropriate choice of $\alpha$ and $\delta$, the factor $e^{2\alpha-2\delta}a$ can be rescaled to either $1$ or $-1$. Hence, we get a twisted subalgebra $\mathfrak{h}_{14}=\lbrace P_++\epsilon P_-,\epsilon=\pm1\rbrace$ given in table~5 in the Appendix.

As another example, consider a twisted algebra of the form $\lbrace K_1^f+\underline{\zeta}W\rbrace$, where $\underline{\zeta}$ is a fermionic parameter. Through the Baker-Campbell-Hausdorff formula (\ref{BCH}), the vector field $Y=K_1^f+\underline{\zeta}W$ transforms through
\be
X=\alpha K_1^f+\beta P_++\underline{\eta}J_++\gamma K_2^f+\delta P_-+\underline{\lambda}J_-+\rho K_0+\sigma C_0+\underline{\tau}W,
\ee
(where $\alpha,\beta,\gamma,\delta,\rho,\sigma\in\Lambda_{even}$ and $\underline{\eta},\underline{\lambda},\underline{\tau}\in\Lambda_{odd}$) to 
\be
e^XYe^{-X}=K_1^f+e^{\rho-\sigma}\underline{\zeta}W-\frac{\beta}{\alpha}(e^{2\alpha}-1)P_+-\frac{1}{\alpha}(e^\alpha-1)\underline{\eta}J_+.\label{BCH3}
\ee
Through a suitable choice of $\beta$ and $\underline{\eta}$, the last two terms of the expression (\ref{BCH3}) can be eliminated, so we obtain the twisted subalgebra $\mathfrak{h}_{32}=\lbrace K_1^f+\underline{\zeta}W\rbrace$ given in table~5 in the Appendix. Continuing the classification in a similar way, involving twisted and non-twisted subalgebras according to \cite{Winternitz}, we obtain the list of one-dimensional subalgebras given in table~5 in the Appendix. These representative subalgebras allow us to determine invariant solutions of the bosonic and fermionic SUSY GC equations, (\ref{BGC}) and (\ref{FGC}) respectively, using the SRM. 

For the specific fermionic case where $f$ is constant (i.e. the SUSY GC equations (\ref{FGC0})), the one-dimensional subalgebras of the resulting Lie symmetry superalgebra (6.6) can be found by taking the limit where the coefficients of $C_0$ tend to zero in the subalgebras listed in table~5 and withdrawing repeated subalgebras, while rescaling appropriately.

\section{Invariant solutions of the SUSY GC equations}\setcounter{equation}{0}
We now make use of the SRM in order to obtain invariant solutions of the bosonic and fermionic SUSY GC equations. For each of the two SUSY GC systems (bosonic and fermionic) we select two representative subalgebras from the corresponding list of subalgebras (table~4 and table~5, respectively) and construct group invariant solutions. For each subalgebra, the invariants and corresponding group orbits are calculated. Next, the unknown superfield functions in $\mathcal{U}$ are expanded in terms of the fermionic symmetry variables $\eta$ and $\sigma$ (involving $\theta^+$ and $\theta^-$, respectively) with coefficients depending on the bosonic symmetry variable $\xi$. Each component $\Upsilon$ of $\mathcal{U}$ is expressed in terms of invariants in the form
\be
\Upsilon=u_0(\xi)+\eta u^+(\xi)+\sigma u^-(\xi)+\eta\sigma u_1(\xi).
\ee
For the sake of simplicity, in what follows, we only consider the case where $u^+=u^-=0$. Substituting these expanded forms of the superfields $\mathcal{U}$ into the SUSY GC equations (\ref{BGC}) and (\ref{FGC}) we reduce these equations to many possible differential subsystems involving even and odd functions. Solving these subsystems, we determine the invariant solutions and provide some geometrical interpretation of the associated surfaces.

For the bosonic SUSY GC equations, we present the following two examples.

\paragraph{}\textbf{Example 1.} In the case of the subalgebra $\mathfrak{g}_{39}=\lbrace P_++\epsilon P_-+aK_0,~\epsilon=\pm1,~a\neq0\rbrace$, the orbit of the group of the bosonic SUSY GC equations (\ref{BGC}) can be parametrized as follows
\be
\ba{ll}
H=e^{-ax_+}h(\xi,\theta^+,\theta^-),&\\
Q^+=e^{ax_+}q^+(\xi,\theta^+,\theta^-),& S^+=s^+(\xi,\theta^+,\theta^-),\\
Q^-=e^{ax_+}q^-(\xi,\theta^+,\theta^-),& S^-=s^-(\xi,\theta^+,\theta^-),\\
R^+=r^+(\xi,\theta^+,\theta^-),& T^+=t^+(\xi,\theta^+,\theta^-),\\
R^-=r^-(\xi,\theta^+,\theta^-),& T^-=t^-(\xi,\theta^+,\theta^-),\\
\phi=2ax_++\varphi(\xi,\theta^+,\theta^-),& f=\psi(\xi),
\ea
\ee
where the functions $H,Q^\pm,R^\pm,S^\pm,T^\pm$ and $\phi$ are expressed in terms of the bosonic symmetry variable $\xi=x_--\epsilon x_+$ and the fermionic symmetry variables $\theta^+$ and $\theta^-$. A corresponding invariant solution is given by
\be
\hspace{-2.5cm}\ba{l}
H=e^{-ax_+}\left[h_0+\theta^+\theta^-2il_0e^\xi\right],\\
Q^+=e^{ax_+}\left[l_0e^{2\xi}+l_1e^\xi+\theta^+\theta^-\left(\frac{1}{2}ie^\xi(ah_0+\epsilon(h_0)_\xi)+l_0e^{2\xi}\varphi_1+l_1e^\xi\varphi_1\right)\right],\\
Q^-=e^{ax_+}\left[\frac{\epsilon l_0}{a\epsilon-1}+l_2e^{(1-a\epsilon)\xi}+\theta^+\theta^-\left(-\frac{1}{2}ie^\xi(h_0)_\xi+\frac{\epsilon l_0}{a\epsilon-1}\varphi_1+l_2e^{(1-a\epsilon)\xi}\varphi_1\right)\right],\\
R^-=b_1\underline{S}^+_0,\quad R^+=b_2\underline{S}^+_0,\quad S^+=\underline{S}_0^+,\quad S^-=\underline{S}_0^+,\quad T^-=b_3\underline{T}^+_0,\quad T^+=b_4\underline{S}^+_0,\\
\phi=2ax_++\xi+\theta^+\theta^-\varphi_1,\qquad f=\psi,\\
l_0=\underline{a}_0\underline{S}_0^+,\qquad\quad l_1=\underline{a}_2\underline{S}_0^+,\qquad l_2=\underline{a}_2\underline{S}_0^+,\qquad\quad h_0=\underline{c}_0\underline{S}_0^+,
\ea\label{Binv1}
\ee
where $h_0$, $\varphi_1$ and $\psi$ are functions of the symmetry variable $\xi=x_--\epsilon x_+$ and where $l_0,l_1,l_2$ and $b_1,b_2,b_3,b_4$ are bosonic constants, while $\underline{S}_0^+,\underline{c}_0$ and $\underline{a}_0,\underline{a}_1,\underline{a}_2$ are fermionic constants.

The first and second fundamental forms of the surface $\mathcal{S}$ associated with (\ref{Binv1}) are given by
\be
\hspace{-2.5cm}\ba{ll}
I=\psi d_+d_-\left[e^{2ax_++\xi}\left(1+\theta^+\theta^-\varphi_1\right)\right],\\
I\hspace{-0.1cm}I=\psi e^{ax_+}\left\lbrace d_+^2\left[l_0e^{2\xi}+l_1e^\xi+\theta^+\theta^-\left(\frac{1}{2}ie^\xi(ah_0+\epsilon(h_0)_\xi)+l_0e^{2\xi}\varphi_1+l_1e^\xi\varphi_1\right)\right]\right.\\
\phantom{I\hspace{-0.1cm}I=\psi e^{ax_+}}+d_+d_-\left[e^\xi\left(h_0+\theta^+\theta^-(2il_0e^\xi+h_0\varphi_1)\right)\right]\\
\left.\phantom{I\hspace{-0.1cm}I=\psi e^{ax_+}}+d_-^2\left[\frac{\epsilon l_0}{a\epsilon-1}+l_2e^{(1-a\epsilon)\xi}+\theta^+\theta^-\left(-\frac{1}{2}ie^\xi(h_0)_\xi+\frac{\epsilon l_0}{a\epsilon-1}\varphi_1+l_2e^{(1-a\epsilon)\xi}\varphi_1\right)\right]\right\rbrace.
\ea
\ee
The Gaussian curvature takes the form
\be
\hspace{-2.6cm}\ba{l}
\mathcal{K}\hspace{-0.1cm}=\hspace{-0.1cm}e^{-2ax_+}\hspace{-0.1cm}\left[h_0^2+\theta^+\theta^-4ih_0l_0e^\xi+4(l_0e^{2\xi}+l_1e^\xi)(\frac{\epsilon l_0}{a\epsilon-1}+l_2e^{(1-a\epsilon)\xi})e^{-2\xi}(1-\theta^+\theta^-2\varphi_1)\right.\\
\phantom{\mathcal{K}=e^{-2ax_+}}+4\theta^+\theta^-(l_0e^{2\xi}+l_1e^\xi)\left(-\frac{1}{2}ie^\xi(h_0)_\xi+\frac{\epsilon l_0}{a\epsilon-1}\varphi_1+l_2e^{(1-a\epsilon)\xi}\varphi_1\right)e^{-2\xi}\\
\phantom{\mathcal{K}=e^{-2ax_+}}\left.+4\theta^+\theta^-(\frac{\epsilon l_0}{a\epsilon-1}+l_2e^{(1-a\epsilon)\xi})\left(\frac{1}{2}ie^\xi(ah_0+\epsilon(h_0)_\xi)+l_0e^{2\xi}\varphi_1+l_1e^\xi\varphi_1\right)e^{-2\xi}\right].
\ea
\ee
The subalgebra of the classical GC equation (\ref{ClaGC}) analogous to $\mathfrak{g}_{39}$ is $L_{1,7}'=\lbrace e_1+\epsilon e_2+ae_0,\epsilon=\pm1,a\neq0\rbrace$, whose corresponding invariant solution is given by
\be
\ba{ll}
H(z,\bar{z})=k_0v(\xi)^{-1/2}e^{a/2(\bar{z}-3z)},& Q(z,\bar{z})=\frac{1}{2}k_0v(\xi)^{1/2}e^{a/2(z+\bar{z})},\\
U(z,\bar{z})=e^{2az}v(\xi),& \bar{Q}(z,\bar{z})=\frac{1}{2}k_0v(\xi)^{1/2}e^{a/2(z+\bar{z})},
\ea
\ee
where the symmetry variable is $\xi=\bar{z}-\epsilon z$ and the function $v$ of $\xi$ satisfies the ODE
\be
v_{\xi\xi}=\frac{(v_\xi)^2}{v}+k_0^2ve^{a\xi}.
\ee
For this classical solution, the Gaussian curvature vanishes, in contrast to the SUSY case.

\paragraph{}\textbf{Example 2.} For the subalgebra $\mathfrak{g}_{76}=\lbrace K_1^b+(a-\frac{1}{2})K_0+\frac{1}{2}C_0,~a\neq\frac{1}{2}\rbrace$ we obtain the following parametrization of the orbit of the group
\be
\ba{ll}
H=(x_+)^{(a-1)/2}h(x_-,\eta,\theta^-),&\\
Q^+=(x_+)^{-(a+2)/2}q^+(x_-,\eta,\theta^-),& S^+=s^+(x_-,\eta,\theta^-),\\
Q^-=(x_+)^{-a/2}q^-(x_-,\eta,\theta^-),& S^-=(x_+)^{-1/2}s^-(x_-,\eta,\theta^-),\\
R^+=(x_+)^{-1/2}r^+(x_-,\eta,\theta^-),& T^+=(x_+)^{1/2}t^+(x_-,\eta,\theta^-),\\
R^-=(x_+)^{-1}r^-(x_-,\eta,\theta^-),& T^-=t^-(x_-,\eta,\theta^-),\\
e^\phi=(x_+)^{-a}\varphi(x_-,\eta,\theta^-),& f=(x_+)^{1/2}\psi(x_-),
\ea
\ee
where the bosonic symmetry variable is $x_-$ and the fermionic symmetry variables are $\eta=(x_+)^{-1/2}\theta^+$ and $\theta^-$. A corresponding invariant solution of the bosonic SUSY GC equations (\ref{BGC}) takes the form
\be
\hspace{-2.5cm}\ba{l}
H=2iB(x_+)^{(a-2)/2}(\rho)_{x_-}\theta^+\theta^-,\\
Q^+=BA(x_-)(x_+)^{-(a+2)/2}\left[1+(x_+)^{-1/2}\theta^+\theta^-G(x_-)\right]\rho(x_-),\\
Q^-=\frac{2B}{a}(x_+)^{-a/2}\left[1+(x_+)^{-1/2}\theta^+\theta^-G(x_-)\right],\\
R^+=(x_+)^{-1/2}l_1\underline{R}_0^+,\hspace{0.3cm} R^-=(x_+)^{-1}l_2\underline{R}_0^-,\hspace{0.3cm} S^+=T^-=\underline{T}_0^-,\hspace{0.3cm} S^-=T^+=0,\\
e^\phi=A(x_-)(x_+)^{-a}(1+(x_+)^{-1/2}\theta^+\theta^-G(x_-)),\qquad f=(x_+)^{1/2}\psi(x_-),
\ea\label{Binv2}
\ee
where $B=\underline{l}_0\underline{R}_0^+\underline{R}_0^-\underline{T}_0^-$ and $l_1,l_2,l_3$ are bosonic constants, while $\underline{l}_0$,  $\underline{R}_0^\pm$ and $\underline{T}_0^-$ are fermionic constants.
Here, $A$, $G$, $\rho$ and $\psi$ are arbitrary bosonic functions of the symmetry variable $x_-$. The function $A$ contains a part in $\Lambda_{body}$ but $\psi$ is a bodiless function.

The corresponding first and second fundamental forms for the surface $\mathcal{S}$ given by (\ref{Binv2}) are
\be
I=\psi d_+d_-\left[A(x_+)^{-(2a+1)/2}\left(1+(x_+)^{-1/2}\theta^+\theta^-G\right)\right],
\ee
and
\be
\hspace{-2cm}\ba{l}
I\hspace{-0.1cm}I=(d_+)^2\left[AB(x_+)^{-(a+2)/2}\rho\left(1+(x_+)^{-1/2}\theta^+\theta^-G\right)\right]\\
\hspace{0.7cm}+2id_+d_-\left[AB(x_+)^{-1}\theta^+\theta^-\rho'\right]+(d_-)^2\left[\frac{2B}{a}(x_+)^{-a/2}\left(1+(x_+)^{-1/2}\theta^+\theta^-G\right)\right].
\ea
\ee
Consequently, the Gaussian curvature $\mathcal{K}$ and the mean curvature $H$ of the associated surface $\mathcal{S}$ are not constant. The Gaussian curvature is given by
\be
\mathcal{K}=\frac{8B}{aA}(x_+)^{a-1}\rho\left(1+(x_+)^{-1/2}\theta^+\theta^-G\right).
\ee
Since $H^2=0$, it follows that the surface $\mathcal{S}$ admits umbilic points along the curve defined by $\mathcal{K}=0$, which lies on the surface $\mathcal{S}$. The subalgebra of the Lie algebra for the classical GC equation (\ref{ClaGC}) analogous to subalgebra $\mathfrak{g}_{76}$ is $L_{1,2}'=\lbrace e_3+ae_0\rbrace$. The corresponding invariant solution is given by
\be
\ba{ll}
H(z,\bar{z})=l_0e^{-a(z+\bar{z})},\qquad Q(z,\bar{z})=k_0e^{a(z+\bar{z})},\\
U(z,\bar{z})=\frac{-2k_0}{l_0}e^{2a(z+\bar{z})},\qquad \bar{Q}(z,\bar{z})=k_0e^{a(z+\bar{z})},\qquad k_0,l_0\in\mathbb{R}.
\ea\label{ClaB2}
\ee
In contrast to the bosonic SUSY case (\ref{Binv2}), the Gaussian curvature $\mathcal{K}$ vanishes for the classical solution (\ref{ClaB2}) associated with the subalgebra $L_{1,2}'$. In both cases however, the mean curvature function $H$ is non-zero.

For the fermionic SUSY GC equations (\ref{FGC}), we present the following two examples.

\paragraph{}\textbf{Example 3.} In the case of the subalgebra $\mathfrak{h}_{124}=\lbrace P_++\epsilon P_-+aK_0,\epsilon=\pm1,a\neq0\rbrace$, the orbit of the corresponding group of the fermionic SUSY GC equations (\ref{FGC}) can be parametrized as follows
\be
\hspace{-2.5cm}\ba{ll}
H=e^{-ax_+}[h_0(\xi)+\theta^+\theta^- h_1(\xi)], & R^+=r^+_0(\xi)+\theta^+\theta^-r^+_1(\xi), \\
Q^+=e^{ax_+}[q_0^+(\xi)+\theta^+\theta^-q_1^+(\xi)], & T^-=r^-_0(\xi)+\theta^+\theta^-r^-_1(\xi), \\
Q^-=e^{ax_+}[q_0^-(\xi)+\theta^+\theta^-q_1^-(\xi)], & \phi=\varphi_0(\xi)+\theta^+\theta^-\varphi_1(\xi)+2ax_+,\quad f=\psi(\xi),
\ea
\ee
where the fermionic functions $H,Q^\pm,R^+$ and $T^-$ are expressed in terms of the bosonic symmetry variable $\xi=x_+-\epsilon x_-$ and the fermionic symmetry variables $\theta^+$ and $\theta^-$, while the bosonic functions $\varphi_0,\varphi_1$ and $\psi$ are expressed in terms of $\xi$ only. A corresponding invariant solution is given by
\be
\hspace{-2.5cm}\ba{l}
H=-2\underline{C}_0^+\underline{C}_0^-e^{-ax_+}\left[\epsilon e^{-\varphi_0}\underline{m}_0^++i\theta^+\theta^-(e^{-\varphi_0}\underline{m}_0^+)_\xi\right],\\
Q^+=-e^{ax_+}\underline{C}_0^+\underline{C}_0^-\left[ \underline{m}_0^++i\theta^+\theta^-((\underline{m}_0^+)_\xi+\epsilon a\underline{m}_0^+)\right],\\
Q^-=e^{ax_+}\underline{C}_0^+\underline{C}_0^-\left[\underline{m}_0^-+i\theta^+\theta^-(\epsilon a\underline{m}_0^-+(\underline{m}_0^-)_\xi)\right],\\
\phi=\varphi_0(\xi)+i\theta^+\theta^-(\varphi_0)_\xi+2ax_+,\quad 
R^+=\underline{C}_0^+,\quad T^-=\underline{C}_0^-,\quad f=\psi(\xi),
\ea\label{Finv1}
\ee
where the fermionic functions $\underline{m}_0^+$, $\underline{m}_0^-$ and the bosonic function $\varphi_0$ of the symmetry variable $\xi$ satisfy the differential constraint
\begin{equation}
[e^{-\varphi_0}(\underline{m}_0^--\epsilon \underline{m}^+_0)]_\xi+\epsilon a\underline{m}_0^-e^{-\varphi_0}=0.
\end{equation}
Here $\psi$ is an arbitrary bosonic function of $\xi$, while $\underline{C}_0^+$ and $\underline{C}_0^-$ are arbitrary fermionic constants.

The first and second fundamental forms of the surface $\mathcal{S}$ associated with the solution (\ref{Finv1}) are given by
\be
\hspace{-2.5cm}\ba{l}
I=\psi e^{\varphi_0+2ax_+}d_+d_-\left[1+\theta^+\theta^-\varphi_1\right],\\
I\hspace{-0.1cm}I=e^{ax_+}\underline{C}_0^+\underline{C}_0^-\psi\left[-2d_+d_-\left(\epsilon \underline{m}_0^++i\theta^+\theta^-\left[(\underline{m}_0^+)_\xi-\epsilon i\varphi_1 \underline{m}_0^+-(\varphi_0)_\xi \underline{m}_0^+\right]\right)\right.\\
\hspace{7mm}\left.+d_+^2\left(\underline{m}_0^++i\theta^+\theta^-\left[(\underline{m}_0^+)_\xi+\epsilon a\underline{m}_0^+\right]\right)+d_-^2\left(\underline{m}_0^-+i\theta^+\theta^-\left[(\underline{m}_0^-)_\xi+\epsilon a\underline{m}_0^-\right]\right)\right].
\ea
\ee
The Gaussian curvature (\ref{SUSYKH}) takes the form
\be
\mathcal{K}=0.
\ee

In particular when $a=0$, which corresponds to the subalgebra $\mathfrak{h}_{14}=\lbrace P_++\epsilon P_-\rbrace$, the orbits of the group of the fermionic SUSY GC equations (\ref{FGC}) can be parametrized in such a way that $H, Q^\pm,R^+$ and $T^-$ are fermionic functions of the bosonic symmetry variable $\xi=x_--\epsilon x_+$, and the fermionic coordinates $\theta^+$ and $\theta^-$ while $\phi$ is a bosonic function of $\xi,\theta^+$ and $\theta^-$, and $\psi$ is a bosonic function of $\xi$ only. Under the assumption that the unknown functions take the form
\be
\ba{ll}
H=h_0(\xi)+\theta^+\theta^- h_1(\xi), & R^+=r_0^+(\xi)+\theta^+\theta^- r_1^+(\xi),\\
Q^\pm=q_0^\pm(\xi)+\theta^+\theta^-q_1^\pm(\xi), & \phi=\varphi_0(\xi)+\theta^+\theta^-\varphi_1(\xi),\\
T^-=t_0^-(\xi)+\theta^+\theta^- t_1^-(\xi), & f=\psi(\xi),
\ea\label{Fpreinv1a}
\ee
the corresponding invariant solution of the fermionic SUSY GC equations (\ref{FGC}) is given by
\be
\hspace{-2.5cm}\ba{l}
H=2\underline{C}_0^-\underline{C}_0^+\underline{l}\left[\int e^{-\varphi_0}d\xi+i\theta^+\theta^-e^{-\varphi_0}\right]+\underline{C},\qquad \epsilon=1,\\
Q^+=\underline{C}_0^-\underline{C}_0^+\underline{l}e^{\varphi_0}\int e^{-\varphi_0}d\xi+\underline{C}_0^-B_0^+e^{\varphi_0}\\
\hspace{2.5cm}+i\theta^+\theta^-\underline{C}_0^-\left[\underline{C}_0^+\underline{l}\left(e^{\varphi_0}(\varphi_0)_\xi\int e^{-\varphi_0}d\xi+1\right)+B_0^+e^{\varphi_0}(\varphi_0)_\xi\right],\\
Q^-=\underline{C}_0^+\underline{C}_0^-\underline{l}e^{\varphi_0}\int e^{-\varphi_0}d\xi+\underline{C}_0^+B_0^-e^{\varphi_0}\\
\hspace{2.5cm}+i\theta^+\theta^-\underline{C}_0^+\left[\underline{C}_0^-\underline{l}\left( e^{\varphi_0}(\varphi_0)_\xi\int e^{-\varphi_0}d\xi+1\right)+B_0^-e^{\varphi_0}(\varphi_0)_\xi\right],\\
R^+=\underline{C}_0^+,\qquad T^-=\underline{C}_0^-,\qquad\phi=\varphi_0(\xi)+i\theta^+\theta^-(\varphi_0(\xi))_\xi,\qquad f=\psi(\xi),
\ea\label{Finv1a}
\ee
where $\varphi_0$ and $\psi$ are bosonic functions of the symmetry variable $\xi=x_--x_+$, while $\underline{C}_0^\pm,\underline{C}$ and $\underline{l}$ are arbitrary fermionic constants and $B_0^\pm$ are bosonic constants satisfying the algebraic constraint
\be
\underline{C}_0^+B_0^-+\underline{C}_0^-B_0^+=0.
\ee
For the solution (\ref{Finv1a}), the tangent vectors are linearly dependent, so the immersion defines curves instead of surfaces.

\paragraph{}\textbf{Example 4.} For the subalgebra $\mathfrak{h}_{35}=\lbrace K_1^f+aK_0+bC_0,a\neq0,b\neq0\rbrace$, we obtain the following parametrization of the orbit of the corresponding supergroup of the fermionic SUSY GC equations (\ref{FGC})
\begin{equation*}\hspace{-2.5cm}\ba{ll}
H=(x_+)^{(a-b)/2}[h_0(x_-)+\eta\theta^-h_1(x_-)], & R^+=(x_+)^{-1/2}[r_0^+(x_-)+\eta\theta^-r_1^+(x_-)], \\
Q^+=(x_+)^{-(a+b+2)/2}[q_0^+(x_-)+\eta\theta^-q_1^+(x_-)], & T^-=t^-_0(x_-)+\eta\theta^-t_1^-(x_-),\\
Q^-=(x_+)^{-(a+b)/2}[q^-_0(x_-)+\eta\theta^-q^-_1(x_-)], & \phi=\varphi_0(x_-)+\eta\theta^-\varphi_1(x_-)-\frac{2a+1}{2}\ln x_+,\\
f=(x_+)^b\psi(x_-), & 
\ea\end{equation*}
where the bosonic symmetry variable is $x_-$ and the fermionic symmetry variables are $\eta=(x_+)^{-1/2}\theta^+$ and $\theta^-$. A corresponding invariant solution of the fermionic SUSY GC equations (\ref{FGC}) has the form
\be
\hspace{-2.5cm}\ba{l}
H=(x_+)^{(a-b)/2}e^{A_0(a-b)x_-/2E_0}\left[\underline{C}+i(x_+)^{-1/2}\theta^+\theta^-(a-b+1)A_0\underline{C}_0^+e^{A_0x_-/2E_0}\right],\\
Q^+=\underline{C}_0^+(x_+)^{-(a+b+2)/2}\left[E_0+(x_+)^{-1/2}\theta^+\theta^-E_1\right],\quad R^+=\underline{C}_0^+(x_+)^{-1/2}, \\
Q^-=A_0\underline{C}_0^+(x_+)^{-(a+b)/2}\left[1+(x_+)^{-1/2}\theta^+\theta^-\frac{E_1}{E_0}\right],\qquad f=(x_+)^b\psi(x_-),\\
T^-=\underline{C}_0^+,\quad\phi=\frac{A_0}{2E_0}(b-a-1)x_-+(x_+)^{-1/2}\theta^+\theta^-\varphi_1(x_-)-\frac{2a+1}{2}\ln x_+,
\ea\label{Finv2}
\ee
where the bosonic function $\varphi_1$ of $x_-$ satisfies the constraint
\be
\underline{C}_0^+\varphi_1=\frac{E_1}{E_0}\underline{C}_0^++i\frac{(a-b)}{4E_0}\underline{C}e^{-A_0x_-/2E_0},
\ee
and where $\psi$ is an arbitrary bosonic function of $x_-$. Here $\underline{C}_0^+$ and $\underline{C}$ are arbitrary fermionic constants while $E_0,E_1$ and $A_0$ are arbitrary bosonic constants.

The first and second fundamental forms of the surface $\mathcal{S}$ (\ref{Finv2}) are given by
\begin{equation*}
\hspace{-2.5cm}\ba{l}
I=(x_+)^{(2b-2a-1)/2}\exp\left(\frac{A_0}{2E_0}(b-a-1)x_-\right)\psi d_+d_-(1+(x_+)^{-1/2}\theta^+\theta^-\varphi_1),\\
I\hspace{-0.1cm}I=(x_+)^{(b-a)/2}\psi\left[\underline{C}_0^+(x_+)^{-1}d_+^2(E_0+(x_+)^{-1/2}\theta^+\theta^- E_1)\right.\\
+A_0\underline{C}_0^+d_-^2(1+(x_+)^{-1/2}\theta^+\theta^-E_1/E_0)\\
+\left.(x_+)^{-1/2}e^{A_0x_-/2E_0}d_+d_-\left(\underline{C}+(x_+)^{-1/2}\theta^+\theta^-\left[i\underline{C}\varphi_1+(a-b+1)A_0\underline{C}_0^+e^{A_0x_-/2E_0}\right]\right)\right].
\ea
\end{equation*}
The Gaussian curvature (\ref{SUSYKH}) takes the form
\be
\mathcal{K}=0.
\ee
Note that in our fermionic SUSY adaptation of the classical geometric interpretation of surfaces in $\mathbb{R}^3$, the surfaces obtained in the two examples are composed of parabolic points.

\section{Conclusions}\setcounter{equation}{0}
In this paper, we formulate bosonic and fermionic SUSY extensions of the GW and GC equations for smooth conformally parametrized surfaces immersed in a Grassmann superspace ($\mathbb{R}^{(2,1\vert2)}$ for the bosonic extension and $\mathbb{R}^{(1,1\vert3)}$ for the fermionic extension). For both SUSY extensions, the GW equations are determined by a moving frame formalism. In the bosonic case, the potential matrices $A_+$ and $A_-$ can be written in subblock form involving both bosonic and fermionic components. In contrast, in the fermionic case, the matrices $A_+$ and $A_-$ are expressed in terms of fermionic quantities only. In the bosonic case, the ZCC for the SUSY GW equations lead to six linearly independent SUSY GC equations in five bosonic functions and six fermionic functions. For the fermionic case, there are four linearly independent SUSY GC equations in two bosonic functions and five fermionic functions. It is interesting to note that, in the latter case, the SUSY GW and SUSY GC equations resemble the form of the classical equations. In the bosonic case, the ZCC involves a diagonal matrix $E$ in addition to the potential matrices $A_+$ and $A_-$. This diagonal matrix becomes the identity matrix in the fermionic case. In both cases, the induced metric involves a bosonic function $f$ of $x_+$ and $x_-$. In the bosonic case, $f$ is bodiless and nilpotent of order $k$, while in the fermionic case the function $f$ may or may not be bodiless. 

For both SUSY GC systems we determine Lie superalgebras of infinitesimal symmetries which generate Lie point symmetry transformations. For both SUSY extensions the symmetry superalgebras include translations in the bosonic independent variables $x_+$ and $x_-$, four dilations and the SUSY operators $J_+$ and $J_-$. The fermionic case contains an additional translation in the direction of the mean curvature $H$. It should be noted that, in the classical case, the Lie point symmetry algebra contains two copies of the Virasoro algebra, whereas such Virasoro algebras do not appear for either the bosonic or the fermionic SUSY GC equations. For both cases, a classification of all one-dimensional subalgebras of the symmetry superalgebra into conjugacy classes is performed. It should be observed that the symmetries of both the classical and the bosonic SUSY GC equations contain a center, whereas the symmetries of the fermionic SUSY GC equations do not. Consequently, the classification lists (by conjugacy classes under the action of the associated supergroup) contain $99$ one-dimensional subalgebras for the bosonic case and 199 one-dimensional subalgebras for the fermionic case. For each of the two SUSY extensions we construct two invariant solutions of the SUSY GC equations and compare them with solutions of the classical GC equations invariant under similar one-dimensional subalgebras. 

The first and second fundamental forms for conformally parametrized surfaces in the superspace are established for both the bosonic and fermionic SUSY extensions of the GC equations. The determinants of the induced metric differs in their signs from the classical case. Also, we establish an analogue of the Bonnet theorem for the SUSY GC equations.

This research could be extended in several directions. It could be beneficial to compute an exhaustive list of all symmetries of the SUSY GC equations and to compare them to the classical case. The computation of such a list would require the development of a computer Lie algebra application capable of handling both even and odd Grassmann variables. Another open problem to be considered is whether all integrable SUSY systems possess non-standard invariants. It could also be worth attempting to establish a SUSY analogue of Noether's theorem in order to study the conservation laws of such SUSY models. Finally, it would be interesting to investigate how integrable characteristics, such as Hamiltonian structure and conserved quantities manifest themselves in surfaces for the SUSY cases. These subjects will be investigated in our future work.

\section*{Acknowledgements}
AMG's work was supported by a research grant from NSERC. SB acknowledges a doctoral fellowship provided by the FQRNT of the Gouvernement du Qu\'ebec. AJH wishes to acknowledge and thank the Mathematical Physics Laboratory of the Centre de Recherches Math\'ematiques for the opportunity to contribute to this research.

\pagebreak

\section*{Appendix. Classification of the one-dimensional subalgebras of the Lie superalgebras (\ref{BsymGC}) and (6.4)}
\begin{table}[h!]
\caption{Classification of the one-dimensional subalgebras of the symmetry superalgebra $\mathfrak{g}$ of the bosonic SUSY GW equations (\ref{BGC}) into conjugacy classes.  Here $\epsilon=\pm1$, the parameters $a,b,m$ are non-zero bosonic constants and $\underline{\mu}$ and $\underline{\nu}$ are non-zero fermionic constants.\vspace{0.5cm}}
\begin{tabular}{|l|l||l|l|}
\hline No. & Subalgebra & No. & Subalgebra \\
\hline $\mathfrak{g}_1$ & $\lbrace K_1^b\rbrace$ & $\mathfrak{g}_2$ & $\lbrace P_+ \rbrace$ \\
\hline $\mathfrak{g}_3$ & $\lbrace \underline{\mu}J_+\rbrace$ & $\mathfrak{g}_4$ & $\lbrace P_++\underline{\mu}J_+\rbrace$ \\
\hline $\mathfrak{g}_5$ & $\lbrace K_2^b\rbrace$ & $\mathfrak{g}_6$ & $\lbrace P_-\rbrace$ \\
\hline $\mathfrak{g}_7$ & $\lbrace \underline{\nu}J_-\rbrace$ & $\mathfrak{g}_8$ & $\lbrace P_-+\underline{\nu}J_-\rbrace$ \\
\hline $\mathfrak{g}_9$ & $\lbrace K_1^b+aK_2^b\rbrace$ & $\mathfrak{g}_{10}$ & $\lbrace K_1^b+\epsilon P_-\rbrace$ \\
\hline $\mathfrak{g}_{11}$ & $\lbrace K_1^b+\underline{\nu}J_-\rbrace$ & $\mathfrak{g}_{12}$ & $\lbrace K_1^b+\epsilon P_-+\underline{\nu}J_-\rbrace$ \\
\hline $\mathfrak{g}_{13}$ & $\lbrace K_2^b+\epsilon P_+\rbrace$ & $\mathfrak{g}_{14}$ & $\lbrace P_++\epsilon P_-\rbrace$ \\
\hline $\mathfrak{g}_{15}$ & $\lbrace P_++\underline{\nu}J_-\rbrace$ & $\mathfrak{g}_{16}$ & $\lbrace P_++\epsilon P_-+\underline{\nu}J_-\rbrace$ \\
\hline $\mathfrak{g}_{17}$ & $\lbrace K_2^b+\underline{\mu}J_+\rbrace$ & $\mathfrak{g}_{18}$ & $\lbrace P_-+\underline{\mu}J_+\rbrace$ \\
\hline $\mathfrak{g}_{19}$ & $\lbrace \underline{\mu}J_++\underline{\nu}J_-\rbrace$ & $\mathfrak{g}_{20}$ & $\lbrace P_-+\underline{\mu}J_++\underline{\nu}J_-\rbrace$ \\
\hline $\mathfrak{g}_{21}$ & $\lbrace K_2^b+\epsilon P_++\underline{\mu}J_+\rbrace$ & $\mathfrak{g}_{22}$ & $\lbrace P_++\epsilon P_-+\underline{\mu}J_+\rbrace$ \\
\hline $\mathfrak{g}_{23}$ & $\lbrace P_++\underline{\mu}J_++\underline{\nu}J_-\rbrace$ & $\mathfrak{g}_{24}$ & $\lbrace P_++\epsilon P_-+\underline{\mu}J_++\underline{\nu}J_-\rbrace$ \\
\hline $\mathfrak{g}_{25}$ & $\lbrace K_0\rbrace$ & $\mathfrak{g}_{26}$ & $\lbrace K_1^b+aK_0\rbrace$ \\
\hline $\mathfrak{g}_{27}$ & $\lbrace K_0+\epsilon P_+\rbrace$ & $\mathfrak{g}_{28}$ & $\lbrace K_0+\underline{\mu}J_+\rbrace$ \\
\hline $\mathfrak{g}_{29}$ & $\lbrace K_0+\epsilon P_++\underline{\mu}J_+\rbrace$ & $\mathfrak{g}_{30}$ & $\lbrace K_2^b+aK_0\rbrace$ \\
\hline $\mathfrak{g}_{31}$ & $\lbrace K_0+\epsilon P_-\rbrace$ & $\mathfrak{g}_{32}$ & $\lbrace K_0+\underline{\nu}J_-\rbrace$ \\
\hline $\mathfrak{g}_{33}$ & $\lbrace K_0+\epsilon P_-+\underline{\nu}J_-\rbrace$ & $\mathfrak{g}_{34}$ & $\lbrace K_1^b+aK_2^b+bK_0\rbrace$ \\
\hline $\mathfrak{g}_{35}$ & $\lbrace K_1^b+aK_0+\epsilon P_-\rbrace$ & $\mathfrak{g}_{36}$ & $\lbrace K_1^b+aK_0+\underline{\nu}J_-\rbrace$ \\
\hline $\mathfrak{g}_{37}$ & $\lbrace K_1^b+aK_0+\epsilon P_-+\underline{\nu}J_-\rbrace$ & $\mathfrak{g}_{38}$ & $\lbrace K_2^b+aK_0+\epsilon P_+\rbrace$ \\
\hline $\mathfrak{g}_{39}$ & $\lbrace K_0+\epsilon_1 P_++\epsilon_2 P_-\rbrace$ & $\mathfrak{g}_{40}$ & $\lbrace K_0+\epsilon P_++\underline{\nu}J_-\rbrace$ \\
\hline $\mathfrak{g}_{41}$ & $\lbrace K_0+\epsilon_1 P_++\epsilon_2 P_-+\underline{\nu}J_-\rbrace$ & $\mathfrak{g}_{42}$ & $\lbrace K_2^b+aK_0+\underline{\mu}J_+\rbrace$ \\
\hline $\mathfrak{g}_{43}$ & $\lbrace K_0+\epsilon P_-+\underline{\mu}J_+\rbrace$ & $\mathfrak{g}_{44}$ & $\lbrace K_0+\underline{\mu}J_++\underline{\nu}J_-\rbrace$ \\
\hline $\mathfrak{g}_{45}$ & $\lbrace K_0+\epsilon P_-+\underline{\mu}J_++\underline{\nu}J_-\rbrace$ & $\mathfrak{g}_{46}$ & $\lbrace K_2^b+aK_0+\epsilon P_++\underline{\mu}J_+\rbrace$ \\
\hline $\mathfrak{g}_{47}$ & $\lbrace K_0+\epsilon_1 P_++\epsilon_2P_-+\underline{\mu}J_+\rbrace$ & $\mathfrak{g}_{48}$ & $\lbrace K_0+\epsilon P_++\underline{\mu}J_++\underline{\nu}J_-\rbrace$ \\
\hline $\mathfrak{g}_{49}$ & $\lbrace K_0+\epsilon_1P_++\epsilon_2P_-+\underline{\mu}J_++\underline{\nu}J_-\rbrace$ & $\mathfrak{g}_{50}$ & $\lbrace C_0\rbrace$ \\
\hline $\mathfrak{g}_{51}$ & $\lbrace K_1^b+aC_0\rbrace$ & $\mathfrak{g}_{52}$ & $\lbrace C_0+\epsilon P_+\rbrace$ \\
\hline $\mathfrak{g}_{53}$ & $\lbrace C_0+\underline{\mu}J_+\rbrace$ & $\mathfrak{g}_{54}$ & $\lbrace C_0+\epsilon P_++\underline{\mu}J_+\rbrace$ \\
\hline $\mathfrak{g}_{55}$ & $\lbrace K_2^b+aC_0\rbrace$ & $\mathfrak{g}_{56}$ & $\lbrace C_0+\epsilon P_-\rbrace$ \\
\hline $\mathfrak{g}_{57}$ & $\lbrace C_0+\underline{\nu}J_-\rbrace$ & $\mathfrak{g}_{58}$ & $\lbrace C_0+\epsilon P_-+\underline{\nu}J_-\rbrace$ \\
\hline $\mathfrak{g}_{59}$ & $\lbrace K_1^b+aK_2^b+bC_0\rbrace$ & $\mathfrak{g}_{60}$ & $\lbrace K_1^b+aC_0+\epsilon P_-\rbrace$ \\
\hline $\mathfrak{g}_{61}$ & $\lbrace K_1^b+aC_0+\underline{\nu}J_-\rbrace$ & $\mathfrak{g}_{62}$ & $\lbrace K_1^b+aC_0+\epsilon P_-+\underline{\nu}J_-\rbrace$ \\
\hline
\end{tabular}
\centering
\end{table}
\setcounter{table}{3}
\begin{table}
\caption{(Continued)}
\begin{tabular}{|l|l||l|l|}
\hline No. & Subalgebra & No. & Subalgebra \\
\hline $\mathfrak{g}_{63}$ & $\lbrace K_2^b+aC_0+\epsilon P_+\rbrace$ & $\mathfrak{g}_{64}$ & $\lbrace C_0+\epsilon_1 P_++\epsilon_2 P_-\rbrace$ \\
\hline $\mathfrak{g}_{65}$ & $\lbrace C_0+\epsilon P_++\underline{\nu}J_-\rbrace$ & $\mathfrak{g}_{66}$ & $\lbrace C_0+\epsilon_1 P_++\epsilon_2 P_-+\underline{\nu}J_-\rbrace$ \\
\hline $\mathfrak{g}_{67}$ & $\lbrace K_2^b+aC_0+\underline{\mu}J_+\rbrace$ & $\mathfrak{g}_{68}$ & $\lbrace C_0+\epsilon P_-+\underline{\mu}J_+\rbrace$ \\
\hline $\mathfrak{g}_{69}$ & $\lbrace C_0+\underline{\mu}J_++\underline{\nu}J_-\rbrace$ & $\mathfrak{g}_{70}$ & $\lbrace C_0+\epsilon P_-+\underline{\mu}J_++\underline{\nu}J_-\rbrace$ \\
\hline $\mathfrak{g}_{71}$ & $\lbrace K_2^b+aC_0+\epsilon P_++\underline{\mu}J_+\rbrace$ & $\mathfrak{g}_{72}$ & $\lbrace C_0+\epsilon_1 P_++\epsilon_2P_-+\underline{\mu}J_+\rbrace$ \\
\hline $\mathfrak{g}_{73}$ & $\lbrace C_0+\epsilon P_++\underline{\mu}J_++\underline{\nu}J_-\rbrace$ & $\mathfrak{g}_{74}$ & $\lbrace C_0+\epsilon_1P_++\epsilon_2P_-+\underline{\mu}J_++\underline{\nu}J_-\rbrace$ \\
\hline $\mathfrak{g}_{75}$ & $\lbrace K_0+mC_0\rbrace$ & $\mathfrak{g}_{76}$ & $\lbrace K_1^b+aK_0+mC_0\rbrace$ \\
\hline $\mathfrak{g}_{77}$ & $\lbrace K_0+mC_0+\epsilon P_+\rbrace$ & $\mathfrak{g}_{78}$ & $\lbrace K_0+mC_0+\underline{\mu}J_+\rbrace$ \\
\hline $\mathfrak{g}_{79}$ & $\lbrace K_0+mC_0+\epsilon P_++\underline{\mu}J_+\rbrace$ & $\mathfrak{g}_{80}$ & $\lbrace K_2^b+aK_0+mC_0\rbrace$ \\
\hline $\mathfrak{g}_{81}$ & $\lbrace K_0+mC_0+\epsilon P_-\rbrace$ & $\mathfrak{g}_{82}$ & $\lbrace K_0+mC_0+\underline{\nu}J_-\rbrace$ \\
\hline $\mathfrak{g}_{83}$ & $\lbrace K_0+mC_0+\epsilon P_-+\underline{\nu}J_-\rbrace$ & $\mathfrak{g}_{84}$ & $\lbrace K_1^b+aK_2^b+bK_0+mC_0\rbrace$ \\
\hline $\mathfrak{g}_{85}$ & $\lbrace K_1^b+aK_0+mC_0+\epsilon P_-\rbrace$ & $\mathfrak{g}_{86}$ & $\lbrace K_1^b+aK_0+mC_0+\underline{\nu}J_-\rbrace$ \\
\hline $\mathfrak{g}_{87}$ & $\lbrace K_1^b+aK_0+mC_0+\epsilon P_-+\underline{\nu}J_-\rbrace$ & $\mathfrak{g}_{88}$ & $\lbrace K_2^b+aK_0+mC_0+\epsilon P_+\rbrace$ \\
\hline $\mathfrak{g}_{89}$ & $\lbrace K_0+mC_0+\epsilon_1 P_++\epsilon_2 P_-\rbrace$ & $\mathfrak{g}_{90}$ & $\lbrace K_0+mC_0+\epsilon P_++\underline{\nu}J_-\rbrace$ \\
\hline $\mathfrak{g}_{91}$ & $\lbrace K_0+mC_0+\epsilon_1 P_++\epsilon_2 P_-+\underline{\nu}J_-\rbrace$ & $\mathfrak{g}_{92}$ & $\lbrace K_2^b+aK_0+mC_0+\underline{\mu}J_+\rbrace$ \\
\hline $\mathfrak{g}_{93}$ & $\lbrace K_0+mC_0+\epsilon P_-+\underline{\mu}J_+\rbrace$ & $\mathfrak{g}_{94}$ & $\lbrace K_0+mC_0+\underline{\mu}J_++\underline{\nu}J_-\rbrace$ \\
\hline $\mathfrak{g}_{95}$ & $\lbrace K_0+mC_0+\epsilon P_-+\underline{\mu}J_++\underline{\nu}J_-\rbrace$ & $\mathfrak{g}_{96}$ & $\lbrace K_2^b+aK_0+mC_0+\epsilon P_++\underline{\mu}J_+\rbrace$ \\
\hline $\mathfrak{g}_{97}$ & $\lbrace K_0+mC_0+\epsilon_1 P_++\epsilon_2P_-+\underline{\mu}J_+\rbrace$ & $\mathfrak{g}_{98}$ & $\lbrace K_0+mC_0+\epsilon P_++\underline{\mu}J_++\underline{\nu}J_-\rbrace$ \\
\hline $\mathfrak{g}_{99}$ & $\lbrace K_0+mC_0+\epsilon_1P_++\epsilon_2P_-+\underline{\mu}J_++\underline{\nu}J_-\rbrace$ &  & \\
\hline
\end{tabular}
\centering
\end{table}

\begin{table}[h!]
\caption{Classification of the one-dimensional subalgebras of the symmetry superalgebra $\mathfrak{h}$ of the equations (\ref{FGC}) into conjugacy classes.  Here $\epsilon=\pm1$, the parameters $a,b$ are non-zero bosonic constants, $\underline{\mu},\underline{\nu},\underline{\zeta}$ are non-zero fermionic constants.}
\begin{tabular}{|l|l||l|l|}
\hline No. & Subalgebra & No. & Subalgebra \\
\hline $\mathfrak{h}_1$ & $\lbrace K_1^f\rbrace$ & $\mathfrak{h}_2$ & $\lbrace P_+\rbrace$ \\
\hline $\mathfrak{h}_3$ & $\lbrace \underline{\mu}J_+\rbrace$ & $\mathfrak{h}_4$ & $\lbrace P_++\underline{\mu}J_+\rbrace$ \\
\hline $\mathfrak{h}_5$ & $\lbrace K_2^f\rbrace$ & $\mathfrak{h}_6$ & $\lbrace P_-\rbrace$ \\
\hline $\mathfrak{h}_7$ & $\lbrace \underline{\nu}J_-\rbrace$ & $\mathfrak{h}_8$ & $\lbrace P_-+\underline{\nu}J_-\rbrace$ \\
\hline $\mathfrak{h}_9$ & $\lbrace K_1^f+aK_2^f\rbrace$ & $\mathfrak{h}_{10}$ & $\lbrace K_1^f+\epsilon P_-\rbrace$ \\
\hline $\mathfrak{h}_{11}$ & $\lbrace K_1^f+\underline{\nu}J_-\rbrace$ & $\mathfrak{h}_{12}$ & $\lbrace K_1^f+\epsilon P_-+\underline{\nu}J_-\rbrace$ \\
\hline $\mathfrak{h}_{13}$ & $\lbrace K_2^f+\epsilon P_+\rbrace$ & $\mathfrak{h}_{14}$ & $\lbrace P_++\epsilon P_-\rbrace$ \\
\hline $\mathfrak{h}_{15}$ & $\lbrace P_++\underline{\nu}J_-\rbrace$ & $\mathfrak{h}_{16}$ & $\lbrace P_++\epsilon P_-+\underline{\nu}J_-\rbrace$ \\
\hline $\mathfrak{h}_{17}$ & $\lbrace K_2^f+\underline{\mu}J_+\rbrace$ & $\mathfrak{h}_{18}$ & $\lbrace P_-+\underline{\mu}J_+\rbrace$ \\
\hline $\mathfrak{h}_{19}$ & $\lbrace \underline{\mu}J_++\underline{\nu}J_-\rbrace$ & $\mathfrak{h}_{20}$ & $\lbrace P_-+\underline{\mu}J_++\underline{\nu}J_-\rbrace$ \\
\hline $\mathfrak{h}_{21}$ & $\lbrace K_2^f+\epsilon P_++\underline{\mu}J_+\rbrace$ & $\mathfrak{h}_{22}$ & $\lbrace P_++\epsilon P_-+\underline{\mu}J_+\rbrace$ \\
\hline $\mathfrak{h}_{23}$ & $\lbrace P_++\underline{\mu}J_++\underline{\nu}J_-\rbrace$ & $\mathfrak{h}_{24}$ & $\lbrace P_++\epsilon P_-+\underline{\mu}J_++\underline{\nu}J_-\rbrace$ \\
\hline $\mathfrak{h}_{25}$ & $\lbrace \underline{\zeta}W\rbrace$ & $\mathfrak{h}_{26}$ & $\lbrace K_0\rbrace$ \\
\hline
\end{tabular}
\centering
\end{table}
\setcounter{table}{4}
\begin{table}
\caption{(Continued)}
\begin{tabular}{|l|l||l|l|}
\hline No. & Subalgebra & No. & Subalgebra \\
\hline $\mathfrak{h}_{27}$ & $\lbrace C_0\rbrace$ & $\mathfrak{h}_{28}$ & $\lbrace K_0+aC_0\rbrace$ \\
\hline $\mathfrak{h}_{29}$ & $\lbrace K_0+\underline{\zeta}W\rbrace$ & $\mathfrak{h}_{30}$ & $\lbrace C_0+\underline{\zeta}W\rbrace$ \\
\hline $\mathfrak{h}_{31}$ & $\lbrace K_0+aC_0+\underline{\zeta}W\rbrace$ & $\mathfrak{h}_{32}$ & $\lbrace K_1^f+\underline{\zeta}W\rbrace$ \\
\hline $\mathfrak{h}_{33}$ & $\lbrace K_1^f+aK_0\rbrace$ & $\mathfrak{h}_{34}$ & $\lbrace K_1^f+aC_0\rbrace$ \\
\hline $\mathfrak{h}_{35}$ & $\lbrace K_1^f+aK_0+bC_0\rbrace$ & $\mathfrak{h}_{36}$ & $\lbrace K_1^f+aK_0+\underline{\zeta}W\rbrace$ \\
\hline $\mathfrak{h}_{37}$ & $\lbrace K_1^f+aC_0+\underline{\zeta}W\rbrace$ & $\mathfrak{h}_{38}$ & $\lbrace K_1^f+aK_0+bC_0+\underline{\zeta}W\rbrace$ \\
\hline $\mathfrak{h}_{39}$ & $\lbrace P_++\underline{\zeta}W\rbrace$ & $\mathfrak{h}_{40}$ & $\lbrace K_0+\epsilon P_+\rbrace$ \\
\hline $\mathfrak{h}_{41}$ & $\lbrace C_0+\epsilon P_+\rbrace$ & $\mathfrak{h}_{42}$ & $\lbrace K_0+aC_0+\epsilon P_+\rbrace$ \\
\hline $\mathfrak{h}_{43}$ & $\lbrace K_0+\epsilon P_++\underline{\zeta}W\rbrace$ & $\mathfrak{h}_{44}$ & $\lbrace C_0+\epsilon P_++\underline{\zeta}W\rbrace$ \\
\hline $\mathfrak{h}_{45}$ & $\lbrace K_0+aC_0+\epsilon P_++\underline{\zeta}W\rbrace$ & $\mathfrak{h}_{46}$ & $\lbrace \underline{\mu}J_++\underline{\zeta}W\rbrace$ \\
\hline $\mathfrak{h}_{47}$ & $\lbrace K_0+\underline{\mu}J_+\rbrace$ & $\mathfrak{h}_{48}$ & $\lbrace C_0+\underline{\mu}J_+\rbrace$ \\
\hline $\mathfrak{h}_{49}$ & $\lbrace K_0+aC_0+\underline{\mu}J_+\rbrace$ & $\mathfrak{h}_{50}$ & $\lbrace K_0+\underline{\mu}J_++\underline{\zeta}W\rbrace$ \\
\hline $\mathfrak{h}_{51}$ & $\lbrace C_0+\underline{\mu}J_++\underline{\zeta}W\rbrace$ & $\mathfrak{h}_{52}$ & $\lbrace K_0+aC_0+\underline{\mu}J_++\underline{\zeta}W\rbrace$ \\
\hline $\mathfrak{h}_{53}$ & $\lbrace P_++\underline{\mu}J_++\underline{\zeta}W\rbrace$ & $\mathfrak{h}_{54}$ & $\lbrace K_0+\epsilon P_++\underline{\mu}J_+\rbrace$ \\
\hline $\mathfrak{h}_{55}$ & $\lbrace C_0+\epsilon P_++\underline{\mu}J_+\rbrace$ & $\mathfrak{h}_{56}$ & $\lbrace K_0+aC_0+\epsilon P_++\underline{\mu}J_+\rbrace$ \\
\hline $\mathfrak{h}_{57}$ & $\lbrace K_0+\epsilon P_++\underline{\mu}J_++\underline{\zeta}W\rbrace$ & $\mathfrak{h}_{58}$ & $\lbrace C_0+\epsilon P_++\underline{\mu}J_++\underline{\zeta}W\rbrace$ \\
\hline $\mathfrak{h}_{59}$ & $\lbrace K_0+aC_0+\epsilon P_++\underline{\mu}J_++\underline{\zeta}W\rbrace$ & $\mathfrak{h}_{60}$ & $\lbrace K_2^f+\underline{\zeta}W\rbrace$ \\
\hline $\mathfrak{h}_{61}$ & $\lbrace K_2^f+aK_0\rbrace$ & $\mathfrak{h}_{62}$ & $\lbrace K_2^f+aC_0\rbrace$ \\
\hline $\mathfrak{h}_{63}$ & $\lbrace K_2^f+aK_0+bC_0\rbrace$ & $\mathfrak{h}_{64}$ & $\lbrace K_2^f+aK_0+\underline{\zeta}W\rbrace$ \\
\hline $\mathfrak{h}_{65}$ & $\lbrace K_2^f+aC_0+\underline{\zeta}W\rbrace$ & $\mathfrak{h}_{66}$ & $\lbrace K_2^f+aK_0+bC_0+\underline{\zeta}W\rbrace$ \\
\hline $\mathfrak{h}_{67}$ & $\lbrace P_-+\underline{\zeta}W\rbrace$ & $\mathfrak{h}_{68}$ & $\lbrace K_0+\epsilon P_-\rbrace$ \\
\hline $\mathfrak{h}_{69}$ & $\lbrace C_0+\epsilon P_-\rbrace$ & $\mathfrak{h}_{70}$ & $\lbrace K_0+aC_0+\epsilon P_-\rbrace$ \\
\hline $\mathfrak{h}_{71}$ & $\lbrace K_0+\epsilon P_-+\underline{\zeta}W\rbrace$ & $\mathfrak{h}_{72}$ & $\lbrace C_0+\epsilon P_-+\underline{\zeta}W\rbrace$ \\
\hline $\mathfrak{h}_{73}$ & $\lbrace K_0+aC_0+\epsilon P_-+\underline{\zeta}W\rbrace$ & $\mathfrak{h}_{74}$ & $\lbrace \underline{\nu}J_-+\underline{\zeta}W\rbrace$ \\
\hline $\mathfrak{h}_{75}$ & $\lbrace K_0+\underline{\nu}J_-\rbrace$ & $\mathfrak{h}_{76}$ & $\lbrace C_0+\underline{\nu}J_-\rbrace$ \\
\hline $\mathfrak{h}_{77}$ & $\lbrace K_0+aC_0+\underline{\nu}J_-\rbrace$ & $\mathfrak{h}_{78}$ & $\lbrace K_0+\underline{\nu}J_-+\underline{\zeta}W\rbrace$ \\
\hline $\mathfrak{h}_{79}$ & $\lbrace C_0+\underline{\nu}J_-+\underline{\zeta}W\rbrace$ & $\mathfrak{h}_{80}$ & $\lbrace K_0+aC_0+\underline{\nu}J_-+\underline{\zeta}W\rbrace$ \\
\hline $\mathfrak{h}_{81}$ & $\lbrace P_-+\underline{\nu}J_-+\underline{\zeta}W\rbrace$ & $\mathfrak{h}_{82}$ & $\lbrace K_0+\epsilon P_-+\underline{\nu}J_-\rbrace$ \\
\hline $\mathfrak{h}_{83}$ & $\lbrace C_0+\epsilon P_-+\underline{\nu}J_-\rbrace$ & $\mathfrak{h}_{84}$ & $\lbrace K_0+aC_0+\epsilon P_-+\underline{\nu}J_-\rbrace$ \\
\hline $\mathfrak{h}_{85}$ & $\lbrace K_0+\epsilon P_-+\underline{\nu}J_-+\underline{\zeta}W\rbrace$ & $\mathfrak{h}_{86}$ & $\lbrace C_0+\epsilon P_-+\underline{\nu}J_-+\underline{\zeta}W\rbrace$ \\
\hline $\mathfrak{h}_{87}$ & $\lbrace K_0+aC_0+\epsilon P_-+\underline{\nu}J_-+\underline{\zeta}W\rbrace$ & $\mathfrak{h}_{88}$ & $\lbrace K_1^f+aK_2^f+\underline{\zeta}W\rbrace$ \\
\hline $\mathfrak{h}_{89}$ & $\lbrace K_0+aK_1^f+bK_2^f\rbrace$ & $\mathfrak{h}_{90}$ & $\lbrace C_0+aK_1^f+bK_2^f\rbrace$ \\
\hline $\mathfrak{h}_{91}$ & $\lbrace K_0+aC_0+bK_1^f+cK_2^f\rbrace$ & $\mathfrak{h}_{92}$ & $\lbrace K_0+aK_1^f+bK_2^f+\underline{\zeta}W\rbrace$ \\
\hline $\mathfrak{h}_{93}$ & $\lbrace C_0+aK_1^f+bK_2^f+\underline{\zeta}W\rbrace$ & $\mathfrak{h}_{94}$ & $\lbrace K_0+aC_0+bK_1^f+cK_2^f+\underline{\zeta}W\rbrace$ \\
\hline $\mathfrak{h}_{95}$ & $\lbrace K_1^f+\epsilon P_-+\underline{\zeta}W\rbrace$ & $\mathfrak{h}_{96}$ & $\lbrace K_0+aK_1^f+\epsilon P_-\rbrace$ \\
\hline $\mathfrak{h}_{97}$ & $\lbrace C_0+aK_1^f+\epsilon P_-\rbrace$ & $\mathfrak{h}_{98}$ & $\lbrace K_0+aC_0+bK_1^f+\epsilon P_-\rbrace$ \\
\hline $\mathfrak{h}_{99}$ & $\lbrace K_0+aK_1^f+\epsilon P_-+\underline{\zeta}W\rbrace$ & $\mathfrak{h}_{100}$ & $\lbrace C_0+aK_1^f+\epsilon P_-+\underline{\zeta}W\rbrace$ \\
\hline
\end{tabular}
\centering
\end{table}
\setcounter{table}{4}
\begin{table}
\caption{(Continued)}
\begin{tabular}{|l|l||l|l|}
\hline No. & Subalgebra & No. & Subalgebra \\
\hline $\mathfrak{h}_{101}$ & $\lbrace K_0+aC_0+bK_1^f+\epsilon P_-+\underline{\zeta}W\rbrace$ & $\mathfrak{h}_{102}$ & $\lbrace K_1^f+\underline{\nu}J_-+\underline{\zeta}W\rbrace$ \\
\hline $\mathfrak{h}_{103}$ & $\lbrace K_0+aK_1^f+\underline{\nu}J_-\rbrace$ & $\mathfrak{h}_{104}$ & $\lbrace C_0+aK_1^f+\underline{\nu}J_-\rbrace$ \\
\hline $\mathfrak{h}_{105}$ & $\lbrace K_0+aC_0+bK_1^f+\underline{\nu}J_-\rbrace$ & $\mathfrak{h}_{106}$ & $\lbrace K_0+aK_1^f+\underline{\nu}J_-+\underline{\zeta}W\rbrace$ \\
\hline $\mathfrak{h}_{107}$ & $\lbrace C_0+aK_1^f+\underline{\nu}J_-+\underline{\zeta}W\rbrace$ & $\mathfrak{h}_{108}$ & $\lbrace K_0+aC_0+bK_1^f+\underline{\nu}J_-+\underline{\zeta}W\rbrace$ \\
\hline $\mathfrak{h}_{109}$ & $\lbrace K_1^f+\epsilon P_-+\underline{\nu}J_-+\underline{\zeta}W\rbrace$ & $\mathfrak{h}_{110}$ & $\lbrace K_0+aK_1^f+\epsilon P_-+\underline{\nu}J_-\rbrace$ \\
\hline $\mathfrak{h}_{111}$ & $\lbrace C_0+aK_1^f+\epsilon P_-+\underline{\nu}J_-\rbrace$ & $\mathfrak{h}_{112}$ & $\lbrace K_0+aC_0+bK_1^f+\epsilon P_-+\underline{\nu}J_-\rbrace$ \\
\hline $\mathfrak{h}_{113}$ & $\lbrace K_0+aK_1^f+\epsilon P_-+\underline{\nu}J_-+\underline{\zeta}W\rbrace$ & $\mathfrak{h}_{114}$ & $\lbrace C_0+aK_1^f+\epsilon P_-+\underline{\nu}J_-+\underline{\zeta}W\rbrace$ \\
\hline $\mathfrak{h}_{115}$ & $\lbrace K_0+aC_0+bK_1^f+\epsilon P_-+\underline{\nu}J_-+\underline{\zeta}W\rbrace$ & $\mathfrak{h}_{116}$ & $\lbrace K_2^f+\epsilon P_++\underline{\zeta}W\rbrace$ \\
\hline $\mathfrak{h}_{117}$ & $\lbrace K_0+aK_2^f+\epsilon P_+\rbrace$ & $\mathfrak{h}_{118}$ & $\lbrace C_0+aK_2^f+\epsilon P_+\rbrace$ \\
\hline $\mathfrak{h}_{119}$ & $\lbrace K_0+aC_0+bK_2^f+\epsilon P_+\rbrace$ & $\mathfrak{h}_{120}$ & $\lbrace K_0+aK_2^f+\epsilon P_++\underline{\zeta}W\rbrace$ \\
\hline $\mathfrak{h}_{121}$ & $\lbrace C_0+aK_2^f+\epsilon P_++\underline{\zeta}W\rbrace$ & $\mathfrak{h}_{122}$ & $\lbrace K_0+aC_0+bK_2^f+\epsilon P_++\underline{\zeta}W\rbrace$ \\
\hline $\mathfrak{h}_{123}$ & $\lbrace P_++\epsilon P_-+\underline{\zeta}W\rbrace$ & $\mathfrak{h}_{124}$ & $\lbrace P_++\epsilon P_-+aK_0\rbrace$ \\
\hline $\mathfrak{h}_{125}$ & $\lbrace P_++\epsilon P_-+aC_0\rbrace$ & $\mathfrak{h}_{126}$ & $\lbrace P_++\epsilon P_-+aK_0+bC_0\rbrace$ \\
\hline $\mathfrak{h}_{127}$ & $\lbrace P_++\epsilon P_-+aK_0+\underline{\zeta}W\rbrace$ & $\mathfrak{h}_{128}$ & $\lbrace P_++\epsilon P_-+aC_0+\underline{\zeta}W\rbrace$ \\
\hline $\mathfrak{h}_{129}$ & $\lbrace P_++\epsilon P_-+aK_0+bC_0\underline{\zeta}W\rbrace$ & $\mathfrak{h}_{130}$ & $\lbrace  P_++\underline{\nu}J_-+\underline{\zeta}W\rbrace$ \\
\hline $\mathfrak{h}_{131}$ & $\lbrace K_0+\epsilon P_++\underline{\nu}J_-\rbrace$ & $\mathfrak{h}_{132}$ & $\lbrace C_0+\epsilon P_++\underline{\nu}J_-\rbrace$ \\
\hline $\mathfrak{h}_{133}$ & $\lbrace K_0+aC_0+\epsilon P_++\underline{\nu}J_-\rbrace$ & $\mathfrak{h}_{134}$ & $\lbrace K_0+\epsilon P_++\underline{\nu}J_-+\underline{\zeta}W\rbrace$ \\
\hline $\mathfrak{h}_{135}$ & $\lbrace C_0+\epsilon P_++\underline{\nu}J_-+\underline{\zeta}W\rbrace$ & $\mathfrak{h}_{136}$ & $\lbrace K_0+aC_0+\epsilon P_++\underline{\nu}J_-+\underline{\zeta}W\rbrace$ \\
\hline $\mathfrak{h}_{137}$ & $\lbrace P_++\epsilon P_-+\underline{\nu}J_-+\underline{\zeta}W\rbrace$ & $\mathfrak{h}_{138}$ & $\lbrace P_++\epsilon P_-+aK_0+\underline{\nu}J_-\rbrace$ \\
\hline $\mathfrak{h}_{139}$ & $\lbrace P_++\epsilon P_-+aC_0+\underline{\nu}J_-\rbrace$ & $\mathfrak{h}_{140}$ & $\lbrace P_++\epsilon P_-+aK_0+bC_0+\underline{\nu}J_-\rbrace$ \\
\hline $\mathfrak{h}_{141}$ & $\lbrace P_++\epsilon P_-+aK_0+\underline{\nu}J_-+\underline{\zeta}W\rbrace$ & $\mathfrak{h}_{142}$ & $\lbrace P_++\epsilon P_-+aC_0+\underline{\nu}J_-+\underline{\zeta}W\rbrace$ \\
\hline $\mathfrak{h}_{143}$ & $\lbrace P_++\epsilon P_-+aK_0+bC_0+\underline{\nu}J_-+\underline{\zeta}W\rbrace$ & $\mathfrak{h}_{144}$ & $\lbrace K_2^f+\underline{\mu}J_++\underline{\zeta}W\rbrace$ \\
\hline $\mathfrak{h}_{145}$ & $\lbrace K_0+aK_2^f+\underline{\mu}J_+\rbrace$ & $\mathfrak{h}_{146}$ & $\lbrace C_0+aK_2^f+\underline{\mu}J_+\rbrace$ \\
\hline $\mathfrak{h}_{147}$ & $\lbrace K_0+aC_0+bK_2^f+\underline{\mu}J_+\rbrace$ & $\mathfrak{h}_{148}$ & $\lbrace K_0+aK_2^f+\underline{\mu}J_++\underline{\zeta}W\rbrace$ \\
\hline $\mathfrak{h}_{149}$ & $\lbrace C_0+aK_2^f+\underline{\mu}J_++\underline{\zeta}W\rbrace$ & $\mathfrak{h}_{150}$ & $\lbrace K_0+aC_0+bK_2^f+\underline{\mu}J_++\underline{\zeta}W\rbrace$ \\
\hline $\mathfrak{h}_{151}$ & $\lbrace P_-+\underline{\mu}J_++\underline{\zeta}W\rbrace$ & $\mathfrak{h}_{152}$ & $\lbrace K_0+\epsilon P_-+\underline{\mu}J_+\rbrace$ \\
\hline $\mathfrak{h}_{153}$ & $\lbrace C_0+\epsilon P_-+\underline{\mu}J_+\rbrace$ & $\mathfrak{h}_{154}$ & $\lbrace K_0+aC_0+\epsilon P_-+\underline{\mu}J_+\rbrace$ \\
\hline $\mathfrak{h}_{155}$ & $\lbrace K_0+\epsilon P_-+\underline{\mu}J_++\underline{\zeta}W\rbrace$ & $\mathfrak{h}_{156}$ & $\lbrace C_0+\epsilon P_-+\underline{\mu}J_++\underline{\zeta}W\rbrace$ \\
\hline $\mathfrak{h}_{157}$ & $\lbrace K_0+aC_0+\epsilon P_-+\underline{\mu}J_++\underline{\zeta}W\rbrace$ & $\mathfrak{h}_{158}$ & $\lbrace \underline{\mu}J_++\underline{\nu}J_-+\underline{\zeta}W\rbrace$ \\
\hline $\mathfrak{h}_{159}$ & $\lbrace K_0+\underline{\mu}J_++\underline{\nu}J_-\rbrace$ & $\mathfrak{h}_{160}$ & $\lbrace C_0+\underline{\mu}J_++\underline{\nu}J_-\rbrace$ \\
\hline $\mathfrak{h}_{161}$ & $\lbrace K_0+aC_0+\underline{\mu}J_++\underline{\nu}J_-\rbrace$ & $\mathfrak{h}_{162}$ & $\lbrace K_0+\underline{\mu}J_++\underline{\nu}J_-+\underline{\zeta}W\rbrace$ \\
\hline $\mathfrak{h}_{163}$ & $\lbrace C_0+\underline{\mu}J_++\underline{\nu}J_-+\underline{\zeta}W\rbrace$ & $\mathfrak{h}_{164}$ & $\lbrace K_0+aC_0+\underline{\mu}J_++\underline{\nu}J_-+\underline{\zeta}W\rbrace$ \\
\hline $\mathfrak{h}_{165}$ & $\lbrace P_-+\underline{\mu}J_++\underline{\nu}J_-+\underline{\zeta}W\rbrace$ & $\mathfrak{h}_{166}$ & $\lbrace K_0+\epsilon P_-+\underline{\mu}J_++\underline{\nu}J_-\rbrace$ \\
\hline $\mathfrak{h}_{167}$ & $\lbrace C_0+\epsilon P_-+\underline{\mu}J_++\underline{\nu}J_-\rbrace$ & $\mathfrak{h}_{168}$ & $\lbrace K_0+aC_0+\epsilon P_-+\underline{\mu}J_++\underline{\nu}J_-\rbrace$ \\
\hline
\end{tabular}
\centering
\end{table}
\setcounter{table}{4}
\begin{table}
\caption{(Continued)}
\begin{tabular}{|l|l||l|l|}
\hline No. & Subalgebra & No. & Subalgebra \\
\hline $\mathfrak{h}_{169}$ & $\lbrace K_0+\epsilon P_-+\underline{\mu}J_++\underline{\nu}J_-+\underline{\zeta}W\rbrace$ & $\mathfrak{h}_{170}$ & $\lbrace C_0+\epsilon P_-+\underline{\mu}J_++\underline{\nu}J_-+\underline{\zeta}W\rbrace$ \\
\hline $\mathfrak{h}_{171}$ & $\lbrace K_0+aC_0+\epsilon P_-+\underline{\mu}J_++\underline{\nu}J_-+\underline{\zeta}W\rbrace$ & $\mathfrak{h}_{172}$ & $\lbrace K_2^f+\epsilon P_++\underline{\mu}J_++\underline{\zeta}W\rbrace$ \\
\hline $\mathfrak{h}_{173}$ & $\lbrace K_0+aK_2^f+\epsilon P_++\underline{\mu}J_+\rbrace$ & $\mathfrak{h}_{174}$ & $\lbrace C_0+aK_2^f+\epsilon P_++\underline{\mu}J_+\rbrace$ \\
\hline $\mathfrak{h}_{175}$ & $\lbrace K_0+aC_0+bK_2^f+\epsilon P_++\underline{\mu}J_+\rbrace$ & $\mathfrak{h}_{176}$ & $\lbrace K_0+aK_2^f+\epsilon P_++\underline{\mu}J_++\underline{\zeta}W\rbrace$ \\
\hline $\mathfrak{h}_{177}$ & $\lbrace C_0+aK_2^f+\epsilon P_++\underline{\mu}J_++\underline{\zeta}W\rbrace$ & $\mathfrak{h}_{178}$ & $\lbrace K_0+aC_0+bK_2^f+\epsilon P_++\underline{\mu}J_++\underline{\zeta}W\rbrace$ \\
\hline $\mathfrak{h}_{179}$ & $\lbrace P_++\epsilon P_-+\underline{\mu}J_++\underline{\zeta}W\rbrace$ & $\mathfrak{h}_{180}$ & $\lbrace P_++\epsilon P_-+aK_0+\underline{\mu}J_+\rbrace$ \\
\hline $\mathfrak{h}_{181}$ & $\lbrace P_++\epsilon P_-+aC_0+\underline{\mu}J_+\rbrace$ & $\mathfrak{h}_{182}$ & $\lbrace P_++\epsilon P_-+aK_0+bC_0+\underline{\mu}J_+\rbrace$ \\
\hline $\mathfrak{h}_{183}$ & $\lbrace P_++\epsilon P_-+aK_0+\underline{\mu}J_++\underline{\zeta}W\rbrace$ & $\mathfrak{h}_{184}$ & $\lbrace P_++\epsilon P_-+aC_0+\underline{\mu}J_++\underline{\zeta}W\rbrace$ \\
\hline $\mathfrak{h}_{185}$ & $\lbrace P_++\epsilon P_-+aK_0+bC_0+\underline{\mu}J_++\underline{\zeta}W\rbrace$ & $\mathfrak{h}_{186}$ & $\lbrace P_++\underline{\mu}J_++\underline{\nu}J_-+\underline{\zeta}W\rbrace$ \\
\hline $\mathfrak{h}_{187}$ & $\lbrace K_0+\epsilon P_++\underline{\mu}J_++\underline{\nu}J_-\rbrace$ & $\mathfrak{h}_{188}$ & $\lbrace C_0+\epsilon P_++\underline{\mu}J_++\underline{\nu}J_-\rbrace$ \\
\hline $\mathfrak{h}_{189}$ & $\lbrace K_0+aC_0+\epsilon P_++\underline{\mu}J_++\underline{\nu}J_-\rbrace$ & $\mathfrak{h}_{190}$ & $\lbrace K_0+\epsilon P_++\underline{\mu}J_++\underline{\nu}J_-+\underline{\zeta}W\rbrace$ \\
\hline $\mathfrak{h}_{191}$ & $\lbrace C_0+\epsilon P_++\underline{\mu}J_++\underline{\nu}J_-+\underline{\zeta}W\rbrace$ & $\mathfrak{h}_{192}$ & $\lbrace K_0+aC_0+\epsilon P_++\underline{\mu}J_++\underline{\nu}J_-+\underline{\zeta}W\rbrace$ \\
\hline $\mathfrak{h}_{193}$ & $\lbrace P_++\epsilon P_-+\underline{\mu}J_++\underline{\nu}J_-+\underline{\zeta}W\rbrace$ & $\mathfrak{h}_{194}$ & $\lbrace P_++\epsilon P_-+aK_0+\underline{\mu}J_++\underline{\nu}J_-\rbrace$ \\
\hline $\mathfrak{h}_{195}$ & $\lbrace P_++\epsilon P_-+aC_0+\underline{\mu}J_++\underline{\nu}J_-\rbrace$ & $\mathfrak{h}_{196}$ & $\lbrace P_++\epsilon P_-+aK_0+bC_0+\underline{\mu}J_++\underline{\nu}J_-\rbrace$ \\
\hline $\mathfrak{h}_{197}$ & $\lbrace P_++\epsilon P_-+aK_0+\underline{\mu}J_++\underline{\nu}J_-+\underline{\zeta}W\rbrace$ & $\mathfrak{h}_{198}$ & $\lbrace P_++\epsilon P_-+aC_0+\underline{\mu}J_++\underline{\nu}J_-+\underline{\zeta}W\rbrace$ \\
\hline $\mathfrak{h}_{199}$ & $\lbrace P_++\epsilon P_-+aK_0+bC_0+\underline{\mu}J_++\underline{\nu}J_-+\underline{\zeta}W\rbrace$ &  & \\
\hline
\end{tabular}
\centering
\end{table}


\section*{References}
\begin{thebibliography}{10}
\expandafter\ifx\csname urlstyle\endcsname\relax
  \providecommand{\doi}[1]{doi:\discretionary{}{}{}#1}\else
  \providecommand{\doi}{doi:\discretionary{}{}{}\begingroup
  \urlstyle{rm}\Url}\fi
  
  
  \bibitem{BJ01}
Bergner Y and Jackiw R 2001 Integrable suspersymmetric fluid mechanics from superstrings, Phys. Lett. A \textbf{284} 146-151.
  
  \bibitem{Crombrugghe}  
de Crombrugghe M and Rittenberg V 1983 Supersymmetric quantum mechanics, Ann. of Phys. \textbf{151} 99-126.
  
  \bibitem{Henkel06}
Henkel M and Unterberger J 2006 Supersymmetric extensions of Schr\"odinger invariance, Nucl. Phys. B \textbf{746}  155-201.
  
  \bibitem{JP00}
Jackiw R and Polychronakos A P 2000 Supersymmetric fluid mechanics, Phys. Rev. D \textbf{62} 085019.

\bibitem{Das}
Das A and Popowicz Z 2002 Supersymmetric polytropic gas dynamics, Phys. Lett. A \textbf{296} 15-26.

\bibitem{Fatyga}
Fatyga B W, Kostelecky V A and Truax D R 1989 Grassmann-valued fluid dynamics, J. Math. Phys. \textbf{30} 1464-1472.

\bibitem{GH11}
Grundland A M and Hariton A J 2011 Supersymmetric formulation of polytropic gas dynamics and its invariant solutions, J. Math. Phys. \textbf{52} 043501.

\bibitem{Hariton}
Hariton A J 2006 Supersymmetric extension of the scalar Born-Infeld equation, J. Phys. A \textbf{39} 7105-7114.

\bibitem{Jackiw}
Jackiw R 2002 \textit{A Particle Theorist's View of Supersymmetric Non-Abelian, Noncommutative Fluid Mechanics and $d$-branes} (New York, Springer-Verlag).
  
\bibitem{TJZW}
Treiman S, Jackiw R, Zumino B and Witten E 1985 \textit{Current Algebra and Anomalies}, (Princetown, Princetown University Press, NJ/World Scientific, Singapore).
  
\bibitem{Chaichian}
Chaichian M and Kulish P P 1978 On the method of the inverse scattering problem and B\"acklund transformations for supersymmetric equations, Phys. Lett. B \textbf{78} 413-416.

\bibitem{MathieuLabelle}
Labelle P and Mathieu P 1991 A new $N=2$ supersymmetric Korteweg-de Vries equation, J. Math. Phys. \textbf{32} 923-927.

\bibitem{Liu}
Liu Q P and Manas M 1998 Pfaffian solutions for the Manin-Radul-Mathieu SUSY KdV and SUSY sine-Gordon equations, Phys. Lett. B \textbf{436} 306-310.

\bibitem{Mathieu}
Mathieu P 1988 Supersymmetric extension of the Korteweg-de Vries equation, J. Math. Phys. \textbf{29} 2499-2506.

\bibitem{Manin}
Manin Y I and Radul A O 1985 A supersymmetric extension of the Kadomtsev-Petviashvili hierarchy, Commun. Math. Phys. \textbf{98} 65-77.

\bibitem{Tian}
Tian K and Liu Q P 2009 A supersymmetric Sawada-Kotera equation, Phys. Lett. A \textbf{373} 1807-1810.

\bibitem{Aratyn}
Aratyn H, Gomes J F, Ymai L H and Zimmerman AH 2008 A class of soliton solutions for the $N=2$ super mKdV/sinh-Gordon hierarchy, J. Phys. A: Math. Theor. \textbf{41} 312001.

\bibitem{Coleman}
Coleman S 1975 Quantum sine-Gordon equation as the massive Thirring model, Phys. Rev. D \textbf{11} 2088.

\bibitem{Grammaticos}
Grammaticos B, Ramani A and Carstea A S 2001 Bilinearization and soliton solutions of the $N=1$ supersymmetric sine-Gordon equation, J. Phys. A: Math. Gen. \textbf{34} 4881-4886.

\bibitem{Gomes}
Gomes J F, Ymai L H and Zimmerman A H 2009 Permutability of B\"acklund transformation for $N=1$ supersymmetric sinh-Gordon, Phys. Lett. A \textbf{373} 1401-1404.

\bibitem{Siddiq05}
Siddiq M and Hassan M 2005 On the linearization of the super sine-Gordon equation, Europhys. Lett. \textbf{70} 149-154.

\bibitem{Siddiq06}
Siddiq M, Hassan M and Saleem U 2006 On Darboux transformation of the supersymmetric sine-Gordon equation, J. Phys. A \textbf{39} 7313-7318.

\bibitem{Witten2}
Witten E 1984 Non-Abelian bosonization in two dimensions, Commun. Math. Phys. \textbf{92} 455.

\bibitem{GHS09}
Grundland A M, Hariton A J and Snobl L 2009 Invariant solutions of the supersymmetric sine-Gordon equation, J. Phys. A: Math Theor. \textbf{42} 335203.

\bibitem{Matveev}
Matveev V B and Salle M A 1991 \textit{Darboux transformations and solitons} (Berlin, Springer-Verlag).

\bibitem{Bluman}
Bluman G W and Anco S C 2002 \textit{Symmetry and Integration Methods for Differential Equations}, Springer, New York.

\bibitem{Clarkson}
Clarkson P A and Winternitz P 1999 \textit{Symmetry reduction and exact solutions of nonlinear partial differential equations. In The Painlev\'e Property, One Century Later} (New York, Conte R, Ed, Springer-Verlag, pp.597-669).

\bibitem{Olver}
Olver P J 1986 \textit{Applications of Lie Groups to Differential Equations} (New York, Springer-Verlag).

\bibitem{SW}
Sattinger D and Weaver O 1986 \textit{Lie Group and Algebras with Applications to Physics, Geometry and Mechanics}, (New York, Springer-Verlag).


\bibitem{Cornwell}
Cornwell J F 1989 \textit{Group Theory in Physics, Volume 3} (London, Academic Press).

\bibitem{DeWitt}
DeWitt B 1984 \textit{Supermanifolds} (Cambridge, Cambridge University Press).

\bibitem{Freed}
Freed D S 1999 \textit{Five Lectures on Supersymmetry}, AMS, New York.

\bibitem{Kac}
Kac V 2002 \textit{Classification of supersymmetries}, (Beijing, In proceedings of the ICM; ICM: Vol. 1, pp. 319-344).

\bibitem{Varadarajan}
Varadarajan V S 2011 \textit{Reflections on Quanta, Symmetries and Supersymmetries} (New York, Springer).
  
\bibitem{RH90}
Roelofs G H and van der Hyligenber N W 1990 Prolongation structures for supersymmetric equations, J. Phys. A: Math. Gen. \textbf{23} 5117-5130.

\bibitem{Winternitz}
Winternitz P 1993 \textit{Lie groups and solutions of nonlinear partial differential equations. In Integrable System, Quantum Groups and Quantum Field Theories} (Netherlands, Ibort L A and Rodriguez M A, Eds, Kluwer, Dordrecht, pp. 429-495).

\bibitem{Delisle}
Delisle L, Hussin V and Zakrzewski W J 2013 Constant curvature solutions of Grassmannian sigma models : (1) Holomorphic solutions, (2) Non-holomorphic solutions, J. Geom. Phys. \textbf{66} 24-36 and \textbf{71} 1-10.

\bibitem{Sasaki}
Sasaki R 1983 General classical solutions of complex Grassmannian $\mathbb{C}P^{N-1}$ sigma models, Phys. Lett. B \textbf{130} 69-72.

\bibitem{Witten}
Witten E 1977 Supersymmetric form of nonlinear sigma model in two dimensions, Phys. Rev. D \textbf{16} 2991-2994.

\bibitem{Baer}
Baer C 2010 \textit{Elementary Differential Geometry}, (Cambridge, Cambridge University Press).

\bibitem{do Carmo}
do Carmo M P 1992 \textit{Riemannian Geometry}, (Boston, Birkh\"auser).

\bibitem{Kuehnel}
Kuehnel W 2005 \textit{Differential Geometry : Curves - Surfaces - Manifolds}, (New York, AMS).

\bibitem{Pressley}
Pressley A 2012 \textit{Elementary Differential Geometry}, (New York, Springer).

\bibitem{Thorpe}
Thorpe J A 1994 \textit{Elementary Topics in Differential Geometry}, (New York, Springer).

\bibitem{Bob}
Bobenko A I 1994 \textit{Surfaces in terms of 2 by 2 matrices. Old and new integrable cases; in Harmonic Maps and Integrable Systems}. Eds Fordy A P and Wood J C (London, Vieweg).

\bibitem{Patera}
Patera J and Winternitz P 1977 Subalgebras of real three- and four-dimensional Lie algebras, J. Math. Phys. \textbf{18}, 7, 1449-1455.

\bibitem{ConteGrundland}
Conte R and Grundland A M 2015 Reductions of the Gauss-Codazzi equations to the sixth Painlev\'e equation (in preparation).

\bibitem{Berezin}
Berezin F A 1996 \textit{The Method of Second Quantization} (New York, Academic Press).
  
\bibitem{Binetruy}
Binetruy P 2006 \textit{Supersymmetry : Theory, Experiment, and Cosmology}, (Oxford, Oxford University Press).

\bibitem{Dine}
Dine M 2007 \textit{Supersymmetry and String Theory : Beyond the Standard Model}, (Cambridge, Cambridge University Press).

\bibitem{Terning}
Terning J 2009 \textit{Modern Supersymmetry: Dynamics and Duality}, (Oxford, Oxford University Press).

\bibitem{Weinberg}
Weinberg S 2005 \textit{The Quantum Theory of Fields, Vol 3 : Supersymmetry}, (Cambridge, Cambridge University Press).

\bibitem{BerezinMono}
Berezin F A 1987 \textit{Introduction to Superanalysis}, (Ed. Kirillov A A, New York, Springer).

\bibitem{Rogers81}
Rogers A 1981 Super Lie groups: global topology and local structure, J. Math. Phys. \textbf{22} 939-945.

\bibitem{Rogers80}
Rogers A 1980 A global theory of supermanifolds, J. Math. Phys. \textbf{21} 1352-1365.

\bibitem{Bianchi}
Bianchi L 1927 \textit{Lazioni di geometria differenziale}, (Zanichelli N, Bologna).

\bibitem{Bonnet}
Bonnet O 1867 M\'emoire sur la th\'eorie des surfaces applicables, J. Ec. Polyt \textbf{42}, 72-92.

\bibitem{GHS11}
Grundland A M, Hariton A J and Snobl L 2011 Invariant solutions of supersymmetric nonlinear wave equations, J. Phys. A: Math. Theor. \textbf{44} 085204.

\bibitem{GH13}
Grundland A M and Hariton A J 2013 Supersymmetric version of the Euler system and its invariant solutions, Symmetry \textbf{5} 253-270, Doi 10.3390/sym5030253



\end{thebibliography}
\end{document}